\documentclass[11pt,a4paper,final]{article}
\usepackage{amsmath}%
\usepackage{amsthm}%
\usepackage{amstext}%
\usepackage{amssymb}%
\usepackage{showkeys}%
\usepackage{epsfig}%
\usepackage{graphicx}%
\usepackage{multicol}
\usepackage{xcolor}
\usepackage{apacite}
\usepackage{lifecon}
\usepackage[top=1in, bottom=1.25in, left=0.8in, right=0.8in]{geometry}

\newtheorem{thm}{Theorem}

\newtheorem{lem}{Lemma}

\begin{document}

\title{Equity--Linked Life Insurances on Maximum of Several Assets}
\author{Battulga Gankhuu\footnote{
Department of Applied Mathematics, National University of Mongolia, Ulaanbaatar, Mongolia;
E-mail: battulga.g@seas.num.edu.mn}}
\date{}

\maketitle 

\begin{abstract}
Economic variables play important roles in any economic model, and sudden and dramatic changes exist in the financial market and economy. 
For this reason, to price and hedge equity--linked life insurance products, including segregated funds and unit--linked life insurance products on maximum price of several assets, this paper introduces Bayesian Markov--Switching Vector Autoregressive (MS--VAR) process. By assuming that a regime--switching process is generated by a homogeneous Markov process and a residual process follows a heteroscedastic model, we obtain joint distribution of endogenous variables and insured's future lifetime random variable under risk--neutral probability probability measure. Using the distribution function, we obtain net single premiums and hedging formulas of the equity--linked life insurance products. An advantage of our model is it depends on economic variables and is not complicated as compared to previous papers.\\[3ex]

\textbf{Keywords:} Economic variables, segregated fund, unit--linked life insurance, Bayesian MS--VAR process, locally risk--minimizing strategy.
\end{abstract}

\section{Introduction}

The first option pricing formula dates back to classic papers of \citeA{Black73} and \citeA{Merton73}. They implicitly introduced a risk--neutral valuation method to arbitrage pricing. But it was not fully developed and appreciated until the works of \citeA{Harrison79} and \citeA{Harrison81}. The basic idea of the risk--neutral valuation method is that discounted price process of an underlying asset is a martingale under some risk--neutral probability measure. The option price is equal to an expected value, with respect to the risk--neutral probability measure, of discounted option payoff. In this paper, to price equity--linked life insurance products on maximum, we use the risk--neutral valuation method in the presence of economic variables. 

Sudden and dramatic changes in the financial market and economy are caused by events such as wars, market panics, or significant changes in government policies. To model those events, some authors used regime--switching models. The regime--switching model was introduced by seminal works of \citeA{Hamilton89,Hamilton90} (see also books of \citeA{Hamilton94} and \citeA{Krolzig97}) and the model is hidden Markov model with dependencies, see \citeA{Zucchini16}. Markov regime--switching models have been introduced before Hamilton (1989), see, for example,  \citeA{Goldfeld73}, \citeA{Quandt58}, and \citeA{Tong83}. The regime--switching model assumes that a discrete unobservable Markov process generates switches among a finite set of regimes randomly and that each regime is defined by a particular parameter set. The model is a good fit for some financial data and has become popular in financial modeling. In particular, according to \citeA{Hardy01}, for monthly TSE 300 and S\&P 500 index returns, there is an evidence that a two--regime model provides a good fit. Recently,  by using the stochastic dividend discount model with regime--switching,  \citeA{Battulga22b} obtained pricing and hedging formulas for options and equity--linked life insurance products. Also, to price and hedge rainbow option, \citeA{Battulga23a} used the regime--switching model. Further, to model required rate of return on stock, \citeA{Battulga23b} applied a two--regime model. The result of the paper reveals that the regime--switching model is good fit for the required rate of return.

Economic variables play important roles in any economic model. In some existing pricing models for equity-linked insurance products, the underlying asset price is governed by some stochastic process and it has not taken into account economic variables such as GDP, inflation, unemployment rate, and so on. For example, the classical papers of equity--linked life insurance used a geometric Brownian motion to capture underlying asset prices, see \citeA{Aase94}, \citeA{Brennan76}, and \citeA{Moller98}. However, the underlying asset price modeled by geometric Brownian motion is not a realistic assumption when it comes to the pricing of equity--linked life insurance products. In reality, for equity--liked life insurance products, the price process of the asset should depend on some economic variables. 

Classic Vector Autoregressive (VAR) process was proposed by \citeA{Sims80} who criticize large--scale macroeconometric models, which are designed to model inter--dependencies of economic variables. Besides \citeA{Sims80}, there are some other important works on multiple time series modeling, see, e.g., \citeA{Tiao81}, where a class of vector autoregressive moving average models was studied. For the VAR process, a variable in the process is modeled by its past values and the past values of other variables in the process. After the work of \citeA{Sims80}, VARs have been used for macroeconomic forecasting and policy analysis. However, if the number of variables in the system increases or the time lag is chosen high, then too many parameters need to be estimated. This will reduce the degrees of freedom of the model and entails a risk of over--parametrization. 

Therefore, to reduce the number of parameters in a high-dimensional VAR process, \citeA{Litterman79} introduced probability distributions for coefficients that are centered at the desired restrictions but that have a small and nonzero variance. Those probability distributions are known as Minnesota prior in Bayesian VAR (BVAR) literature which is widely used in practice. Due to over--parametrization, the generally accepted result is that forecast of the BVAR model is better than the VAR model estimated by the frequentist technique. The BVAR relies on Monte--Carlo simulation methods. Recently, for Bayesian Markov--Switching VAR process, \citeA{Battulga24g} introduced a new Monte--Carlo simulation method that removes duplication in a regime vector. Also, the author introduced importance sampling method to estimate probability of rare event, which corresponds to endogenous variables. Research works have shown that BVAR is an appropriate tool for modeling large data sets, for example, see \citeA{Banbura10}.

In this paper, to partially fill the gaps mentioned above, we introduced a Bayesian Markov--Switching VAR (MS--VAR) model to value and hedge equity-linked insurance products. Our model offers the following advantages: (i) it tries to mitigate the valuation complexity of life insurance products which depend on regime--switching, see \citeA{Hardy01}, (ii) it considers economic variables thus the model will be more consistent with future economic uncertainty (iii) it introduces regime--switching so that the model takes into account sudden and dramatic changes in the economy and financial market (iv) it adopts a Bayesian procedure to deal with over--parametrization. Our model talks about not only the Bayesian MS--VAR process but also heteroscedastic models for a residual process. The novelty of the paper is that we introduced Bayesian MS--VAR process which is widely used to model economic variables to equity--linked life insurance products. The paper considers non--dividend paying assets. For dividend--paying assets, we refer to \citeA{Battulga22b}. 

The rest of the paper is structured as follows. Section 2 provides a Theorem which is used to price life insurance products. We briefly consider some results from \citeA{Battulga24a} in Section 3, where the author obtained pricing formulas for some frequently used options under Bayesian MS--VAR process. In Section 4, we obtain pricing formulas for unit--linked life insurance products and segregated funds on a maximum of several asset prices. Section 5 provides hedging formulas, which are based on the locally risk--minimizing strategy for the equity--linked life insurance products. Finally, Section 6 concludes the study.

\section{Preliminary Results}

Let $(\Omega,\mathcal{H}_T^x,\mathbb{P})$ be a complete probability space, where $\mathbb{P}$ is a given physical or real--world probability measure. Other elements of the probability space will be defined below. To introduce a regime--switching in option pricing, we assume that $\{s_t\}_{t=1}^T$ is a homogeneous Markov chain with $N$ state and $\mathsf{P}:=\{p_{ij}\}_{i=0,j=1}^N$ is a random transition probability matrix, including an initial probability vector $\{p_{0j}\}_{j=1}^N$. We consider a Bayesian Markov--Switching Vector Autoregressive (MS--VAR($p$)) process of $p$ order, which is given by the following equation 
\begin{equation}\label{07001}
y_t=A_{0,s_t}\psi_t+A_{1,s_t}y_{t-1}+\dots +A_{p,s_t}y_{t-p}+\xi_t,~t=1,\dots,t,
\end{equation}
where $y_t=(y_{1,t},\dots,y_{n,t})'$ is an $(n\times 1)$ endogenous random vector, $\psi_t=(1,\psi_{2,t},\dots,\psi_{k,t})'$ is a $(k\times 1)$ vector of exogenous variables, $\xi_t=(\xi_{1,t},\dots,\xi_{n,t})'$ is an $(n\times 1)$ residual process, $A_{0,s_t}$ is an $(n\times k)$ random coefficient matrix at regime $s_t$ that corresponds to the vector of exogenous variables, for $i=1,\dots,p$, $A_{i,s_t}$ are random $(n\times n)$ coefficient matrices at regime $s_t$ that correspond to $y_{t-1},\dots,y_{t-p}$. Equation \eqref{07001} can be compactly written by
\begin{equation}\label{07002}
y_t=\Pi_{s_t}\mathsf{Y}_{t-1}+\xi_t,~t=1,\dots,t,
\end{equation}
where $\Pi_{s_t}:=[A_{0,s_t}: A_{1,s_t}:\dots:A_{p,s_t}]$ is a random coefficient matrix at regime $s_t$, which consist of all the random coefficient matrices and $\mathsf{Y}_{t-1}:=(\psi_t',y_{t-1}',\dots,y_{t-p}')'$ is a vector, which consist of exogenous variable $\psi_t$ and last $p$ lagged values of the process $y_t$. In the paper, this form of the MS--BVAR process $y_t$ will play a major role than the form, which is given by equation \eqref{07001}.

To introduce heteroskedasticity, for the residual process $\xi_t$, we assume that it has $\xi_t:=\Sigma_t^{1/2}\varepsilon_t \in\mathbb{R}^n$, $t=1,\dots,t$ representation, see \citeA{Lutkepohl05} and \citeA{McNeil05}, where $\Sigma_t^{1/2}:=\Sigma^{1/2}(\bar{s}_t)$ is the Cholesky factor of a positive definite random matrix $\Sigma(\bar{s}_t)\in\mathbb{R}^{n\times n}$, which depends on a regime vector $\bar{s}_t=(s_1,\dots,s_t)'$ and a random coefficient matrix $\Gamma_{s_t}:=[B_{0,s_t}:B_{1,s_t}:\dots:B_{p_*,s_t}]$ but does not depend on the residual process $\xi_t$. Here $B_{0,s_t}$ is an $(n_*\times k_*)$ random matrix, for $i=1,\dots,p_*$, $B_{i,s_t}$ are $(n_*\times n_*)$ random matrices, and $\varepsilon_1,\dots,\varepsilon_T$ is a random sequence of independent identically multivariate normally distributed random vectors with means of 0 and covariance matrices of $n$ dimensional identity matrix $I_n$. Then, in particular, for multivariate GARCH process of $(0,p_*)$ order, dependence of $\Sigma_t$ on $\Gamma_{s_t}$ is given by 
\begin{equation}\label{07003}
\text{vech}\big(\Sigma_t\big)=B_{0,s_t}+\sum_{i=1}^{p_*}B_{i,s_t}\text{vech}(\Sigma_{t-i}),
\end{equation}
where $B_{0,s_t}\in \mathbb{R}^{n(n+1)/2}$ and $B_{i,s_t}\in \mathbb{R}^{[n(n+1)/2]\times [n(n+1)/2]}$ for $i=1,\dots, p_*$ are suitable random vector and matrices and the vech is an operator that stacks elements on and below a main diagonal of a square matrix. 

Let us introduce stacked vectors and matrices: $y:=(y_1',\dots,y_T')'$, $s:=(s_1,\dots,s_T)'$, $\Pi_s:=[\Pi_{s_1}:\dots:\Pi_{s_T}]$, and $\Gamma_s:=[\Gamma_{s_1}:\dots:\Gamma_{s_T}]$. We also assume that the strong white noise process $\{\varepsilon_t\}_{t=1}^T$ is independent of the random coefficient matrices $\Pi_s$ and $\Gamma_s$, random transition matrix $\mathsf{P}$, and regime--switching process $\{s_t\}_{t=1}^T$ conditional on initial information $\mathcal{F}_0:=\sigma(y_{1-p},\dots,y_0,\psi_{1},\dots,\psi_T,\Sigma_{1-q_*},\dots,\Sigma_0)$. Here for a generic random vector $X$, $\sigma(X)$ denotes a $\sigma$--field generated by the random vector $X$, $y_{1-p},\dots,y_0$ are initial values of endogenous variables, $\Sigma_{1-q_*},\dots,\Sigma_0$ are initial values of the random matrix process $\Sigma_t$, and $\psi_1,\dots,\psi_T$ are exogenous variables and they are known at time zero. We further suppose that the transition probability matrix $\mathsf{P}$ is independent of the random coefficient matrices $\Pi_s$ and $\Gamma_s$ given initial information $\mathcal{F}_0$ and regime--switching process $s$.

To ease of notations, for a generic vector $o=(o_1',\dots,o_T')'$, we denote its first $t$ and last $T-t$ sub vectors by $\bar{o}_t$ and $\bar{o}_t^c$, respectively, that is, $\bar{o}_t:=(o_1',\dots,o_t')'$ and $\bar{o}_t^c:=(o_{t+1}',\dots,o_T')'$. We define $\sigma$--fields: for $t=0,\dots,t$, $\mathcal{F}_{t}:=\mathcal{F}_0\vee\sigma(\bar{y}_{t})$, $\mathcal{H}_{t}:=\mathcal{F}_t\vee \sigma(\Pi_s)\vee \sigma(\Gamma_s)\vee \sigma(\mathsf{P})\vee\sigma(s)$ and for $t=1,\dots,t$, $\mathcal{I}_{t-1}=\mathcal{F}_{t-1}\vee\sigma(\Pi_{\bar{s}_t})\vee\sigma(\Gamma_{\bar{s}_t})\vee \sigma(\mathsf{P})\vee \sigma(\bar{s}_t)$, where for generic sigma fields $\mathcal{O}_1,\dots,\mathcal{O}_k$, $\vee_{i=1}^k \mathcal{O}_i $ is the minimal $\sigma$--field containing the $\sigma$--fields $\mathcal{O}_i$, $i=1,\dots,k$ and $\Pi_{\bar{s}_t}:=[\Pi_{s_1}:\dots:\Pi_{s_t}]$ and $\Gamma_{\bar{s}_t}:=[\Gamma_{s_1}:\dots:\Gamma_{s_t}]$. Observe that $\mathcal{F}_{t}\subset \mathcal{H}_{t}$ for $t=0,\dots,t$. The $\sigma$--fields play major roles in the paper. For the first--order Markov chain, a conditional probability that the regime at time $t+1$, $s_{t+1}$ equals some particular value conditional on the past regimes, $\bar{s}_t$, transition probability matrix $\mathsf{P}$, and initial information $\mathcal{F}_0$ depends only through the most recent regime at time $t$, $s_t$, transition probability matrix $\mathsf{P}$, and initial information $\mathcal{F}_0$, that is,
\begin{equation}\label{07004}
p_{s_ts_{t+1}}:=\mathbb{P}[s_{t+1}=s_{t+1}|s_t=s_t,\mathsf{P},\mathcal{F}_0]=\mathbb{P}\big[s_{t+1}=s_{t+1}|\bar{s}_t=\bar{s}_t,\mathsf{P},\mathcal{F}_0\big]
\end{equation} 
for $t=0,\dots,t-1$, where $p_{s_0s_1}:=p_{0s_1}=\mathbb{P}[s_1=s_1|\mathsf{P},\mathcal{F}_0]$ is the initial probability.
A distribution of a residual random vector $\xi:=(\xi_1',\dots,\xi_T')'$ is given by
\begin{equation}\label{07005}
\xi=(\xi_1',\dots,\xi_T')'~|~\mathcal{H}_0\sim \mathcal{N}(0,\Sigma)
\end{equation}
where $\Sigma:=\text{diag}\{\Sigma_1,\dots,\Sigma_T\}$ is a block diagonal covariance matrix. Let $\bar{\Sigma}_t:=\text{diag}\{\Sigma_1,\dots,\Sigma_t\}$ and $\bar{\Sigma}_t^c:=\text{diag}\{\Sigma_{t+1},\dots,\Sigma_T\}$ be partitions, corresponding to random vectors $\bar{\xi}_t$ and $\bar{\xi}_t^c$ of the covariance matrix $\Sigma$. 

To remove duplicates in the random coefficient matrix $(\Pi_s,\Gamma_s)$, for a generic regime--switching vector with length $k$, $o=(o_1,\dots,o_k)'$, we define sets
\begin{equation}\label{08006}
\mathcal{A}_{\bar{o}_t}:=\mathcal{A}_{\bar{o}_{t-1}}\cup\big\{o_t\in \{o_1,\dots,o_k\}\big|o_t\not \in \mathcal{A}_{\bar{o}_{t-1}}\big\},~~~t=1,\dots,k,
\end{equation}
where for $t=1,\dots,k$, $o_t\in \{1,\dots,N\}$ and an initial set is the empty set, i.e., $\mathcal{A}_{\bar{o}_0}=\O$. The final set $\mathcal{A}_o=\mathcal{A}_{\bar{o}_k}$ consists of different regimes in regime vector $o=\bar{o}_k$ and $|\mathcal{A}_o|$ represents a number of different regimes in the regime vector $o$. Let us assume that elements of sets $\mathcal{A}_s$, $\mathcal{A}_{\bar{s}_t}$, and difference sets between the sets $\mathcal{A}_{\bar{s}_t^c}$ and $\mathcal{A}_{\bar{s}_t}$ are given by $\mathcal{A}_s=\{\hat{s}_1,\dots,\hat{s}_{r_{\hat{s}}}\}$, $\mathcal{A}_{\bar{s}_t}=\{\alpha_1,\dots,\alpha_{r_\alpha}\}$, and $\mathcal{A}_{\bar{s}_t^c}\backslash \mathcal{A}_{\bar{s}_t}=\{\delta_1,\dots,\delta_{r_\delta}\}$, respectively, where $r_{\hat{s}}:=|\mathcal{A}_s|$, $r_\alpha:=|\mathcal{A}_{\bar{s}_t}|$, and $r_\delta:=|\mathcal{A}_{\bar{s}_t^c}\backslash \mathcal{A}_{\bar{s}_t}|$ are numbers of elements of the sets, respectively. We introduce the following regime vectors: $\hat{s}:=(\hat{s}_1,\dots,\hat{s}_{r_{\hat{s}}})'$ is an $(r_{\hat{s}}\times 1)$ vector, $\alpha:=(\alpha_1,\dots,\alpha_{r_\alpha})'$ is an $(r_\alpha\times 1)$ vector, and $\delta=(\delta_1,\dots,\delta_{r_\delta})'$ is an $(r_\delta\times 1)$ vector. For the regime vector $a=(a_1,\dots,a_{r_a})' \in\{\hat{s},\alpha,\delta\}$, we also introduce duplication removed random coefficient matrices, whose block matrices are different:  $\Pi_a=[\Pi_{a_1}:\dots:\Pi_{a_{r_a}}]$ is an $(n\times [(np+k)r_a])$ matrix, $\Gamma_a=[\Gamma_{a_1}:\dots:\Gamma_{a_{r_a}}]$ is an $(n_*\times [(n_*p_*+k_*)r_a])$ matrix, and $(\Pi_a,\Gamma_a)$. 

We assume that for given duplication removed regime vector $\hat{s}$ and initial information $\mathcal{F}_0$, the coefficient matrices $(\Pi_{\hat{s}_1},\Gamma_{\hat{s}_1}),\dots,(\Pi_{\hat{s}_{r_{\hat{s}}}},\Gamma_{\hat{s}_{r_{\hat{s}}}})$ are independent under the real probability measure $\mathbb{P}$. Under the assumption, conditional on $\hat{s}$ and $\mathcal{F}_0$, a joint density function of the random coefficient random matrix $(\Pi_{\hat{s}},\Gamma_{\hat{s}})$ is represented by
\begin{equation}\label{08010}
f\big(\Pi_{\hat{s}},\Gamma_{\hat{s}}\big|\hat{s},\mathcal{F}_0\big)=\prod_{t=1}^{r_{\hat{s}}}f\big(\Pi_{\hat{s}_t},\Gamma_{\hat{s}_t}\big|\hat{s}_t,\mathcal{F}_0\big)
\end{equation}
under the real probability measure $\mathbb{P}$, where for a generic random vector $X$, we denote its density function by $f(X)$ under the real probability measure $\mathbb{P}$. Using the regime vectors $\alpha$ and $\delta$, the above joint density function can be written by 
\begin{equation}\label{08011}
f\big(\Pi_{\hat{s}},\Gamma_{\hat{s}}\big|\hat{s},\mathcal{F}_0\big)=
f\big(\Pi_{\alpha},\Gamma_{\alpha}\big|\alpha,\mathcal{F}_0\big)f_*\big(\Pi_{\delta},\Gamma_{\delta}\big|\delta,\mathcal{F}_0\big)
\end{equation}
where the density function $f_*\big(\Pi_{\delta},\Gamma_{\delta}\big|\delta,\mathcal{F}_0\big)$ equals
\begin{equation}\label{08012}
f_*\big(\Pi_\delta,\Gamma_\delta\big|\delta,\mathcal{F}_0\big):=
\begin{cases}
f\big(\Pi_\delta,\Gamma_\delta\big|\delta,\mathcal{F}_0\big),& \text{if}~~~r_\delta\neq 0,\\
1,& \text{if}~~~r_\delta= 0.
\end{cases}
\end{equation}

In order to change from the real probability measure $\mathbb{P}$ to some risk--neutral probability measure $\tilde{\mathbb{P}}$, we define the following state price density process:
\begin{equation}\label{07006}
L_t:=\prod_{m=1}^t\exp\bigg\{\theta_m'\Sigma_m^{-1}\xi_m-\frac{1}{2}\theta_m'\Sigma_m^{-1}\theta_m\bigg\}
\end{equation}
for $t=1,\dots,t$, where $\theta_m\in \mathbb{R}^n$ is $\mathcal{I}_{m-1}$ measurable Girsanov kernel process (see, \citeA{Bjork09}) and is defined below. Then it can be shown that $\{L_t\}_{t=0}^T$ is a martingale with respect to a filtration $\{\mathcal{H}_t\}_{t=0}^T$ and the real probability measure $\mathbb{P}$. Therefore, we have $\mathbb{E}[L_T|\mathcal{H}_{0}]=\mathbb{E}[L_1|\mathcal{H}_0]=1$ and $\mathbb{E}[L_T|\mathcal{F}_0]=\mathbb{E}\big[\mathbb{E}[L_T|\mathcal{H}_0]|\mathcal{F}_0\big]=1$. 

We denote life--age--$x$ person's future lifetime random variable by $T_x$. Let for $t=0,1,\dots,T$, $I_t:=1_{\{T_x\leq t\}}$ is a death indicator process, where we denote for a generic set $A\in \mathcal{H}_T^x$, its indicator function by $1_A$. This process has non--decreasing right--continuous paths, it jumps from zero to one at the moment of death and interprets the history of the death information up to and including time $t$. A $\sigma$--field which is generated by the death indicator process is defined by $\mathcal{T}_t^x:=\sigma\big(1_{\{T_x\leq s\}}:s\in[0,t]\big)$. We suppose that the $\sigma$--field $\mathcal{T}_T^x$ is independent of the $\sigma$--field $\mathcal{H}_T$, and $\mathcal{F}_t^x:=\mathcal{F}_t\vee \mathcal{T}_t^x$ and $\mathcal{H}_t^x:=\mathcal{H}_t\vee\mathcal{T}_t^x$ for $t=0,\dots,t$. Observe that $\mathcal{F}_0^x=\mathcal{F}_0$ and $\mathcal{H}_0^x=\mathcal{H}_0$ since $\mathcal{T}_0^x=\{\O,\Omega\}$. For the death indicator process, it is well--known fact that (see \citeA{McNeil05}) $$\mathbb{P}[T_x>t~|~\mathcal{T}_{t-1}^x]=\mathbb{P}[I_t=0~|~\mathcal{T}_{t-1}^x]=(1-I_{t-1})\frac{{}_tp_x}{{}_{t-1}p_x},$$
where $_tp_x=\mathbb{P}[T_x>t]$ represents a probability that life--age--$x$ attain age $x+t$. Since $\{T_x\leq t-1\}\subset \{T_x\leq t\}$, the fact that $\Delta I_t:=I_t-I_{t-1}=1_{\{t-1<T_x\leq t\}}$ is clear. So for all $t=1,\dots,t$ and any real vector $g=(g_1,\dots,g_T)'\in \mathbb{R}^T$, one can obtain that
\begin{equation}\label{07007}
\mathbb{E}\big[\exp\{g_{t}\Delta I_t\}~|~\mathcal{T}_{t-1}^x\big]=1+(1-I_{t-1})\big(\exp\{g_{t}\}-1\big)\frac{{}_{t-1}p_x-{}_tp_x}{{}_{t-1}p_x}.
\end{equation}
Based on the above equation, we can define the following state price density process
\begin{equation}\label{07008}
K_t:=\prod_{m=1}^t\exp\bigg\{g_{m}\Delta I_{m}-(1-I_{m-1})\ln\bigg(1+\big(\exp\{g_{m}\}-1\big)\frac{{}_{m-1}p_x-{}_mp_x}{{}_{m-1}p_x}\bigg)\bigg\}
\end{equation}
for $t=1,\dots,t$. According to construction, it is obvious that $K_t$ is a martingale with respect to a filtration $\{\mathcal{T}_t^x\}_{t=0}^T$ and the real probability measure $\mathbb{P}$. Therefore, $\mathbb{E}[K_T]=\mathbb{E}[K_1]=1$. 

Since the state price process $K_T$ is independent of the state price process $L_T$ and the initial information $\mathcal{F}_0$ under the real probability measure $\mathbb{P}$, it is obvious that $\mathbb{E}[M_T|\mathcal{F}_0]=\mathbb{E}[L_T|\mathcal{F}_0]\times \mathbb{E}[K_T]=1$, where $M_T:=L_T\times K_T$. As a result, since for all $\omega\in\Omega$, $M_T(\omega)>0$,
\begin{equation}\label{07009}
\tilde{\mathbb{P}}\big[A|\mathcal{F}_0\big]=\int_AM_T(\omega|\mathcal{F}_0)d\mathbb{P}\big[\omega|\mathcal{F}_0\big]~~~ \mbox{for all}~A\in \mathcal{H}_T^x
\end{equation}
becomes a probability measure, which is called the risk--neutral probability measure. 

By introducing the concept of mean--self--financing, \citeA{Follmer86} extended the concept of the complete market into the incomplete market. In this paper, we will work in an incomplete market. For this reason, we consider a variance of the state price density process 
\begin{equation}\label{07010}
\text{Var}\big[M_T|\mathcal{F}_0\big]=\bigg\|\frac{d\tilde{\mathbb{P}}}{d\mathbb{P}}-1\bigg\|_2^2
\end{equation}
where $\|\cdot\|_2$ is a $\mathcal{L}_2$ norm, and a relative entropy of the risk--neutral probability measure $\mathbb{\tilde{P}}$ with respect to the real probability measure $\mathbb{P}$, is defined by
\begin{equation}\label{07011}
I(\mathbb{\tilde{P}},\mathbb{P})=\mathbb{E}\big[M_T\ln(M_T)\big|\mathcal{F}_0\big]=\mathbb{\tilde{E}}\big[\ln(M_T)\big|\mathcal{F}_0\big].
\end{equation}
Their usage and connection with the incomplete market can be found in \citeA{Frittelli00} and \citeA{Schweizer95}. 

We assume that for $t=1,\dots,T$, $\mathcal{I}_{t-1}$ measurable random process $\theta_t\in \mathbb{R}^n$ has the following representation
\begin{equation}\label{07012}
\theta_t=\Delta_{0,t}\psi_t+\Delta_{1,t}y_{t-1}+\dots+\Delta_{p,t}y_{t-p},~~~t=1,\dots,T,
\end{equation}
where $\Delta_{0,t}\in\mathbb{R}^{n\times k}$ and $\Delta_{i,t}\in\mathbb{R}^{n\times n}$, $i = 1,\dots,p$ are $\mathcal{I}_{t-1}$ measurable random coefficient matrices. If we define the following matrix and vectors:
\begin{equation}\label{07013}
\Psi:=\begin{bmatrix}
I_n & 0 & \dots & 0 & \dots & 0 & 0\\
-A_{1,s_2}-\Delta_{1,2} & I_n & \dots & 0 & \dots & 0 & 0\\
\vdots & \vdots & \dots & \vdots & \dots & \vdots & \vdots\\
0 & 0 & \dots & -A_{p-1,s_{T-1}}-\Delta_{p-1,T-1} & \dots & I_n & 0\\
0 & 0 & \dots & -A_{p,s_T}-\Delta_{p,T} & \dots & -A_{1,s_T}-\Delta_{1,T} & I_n
\end{bmatrix},
\end{equation}
and
\begin{equation}\label{07014}
\delta:=\begin{bmatrix}
(A_{0,s_1}+\Delta_{0,1})\psi_1+(A_{1,s_1}+\Delta_{1,1})y_{0}+\dots+(A_{p,s_1}+\Delta_{p,1})y_{1-p}\\ 
(A_{0,s_2}+\Delta_{0,2})\psi_2+(A_{2,s_2}+\Delta_{2,2})y_{0}+\dots+(A_{p,s_2}+\Delta_{p,2})y_{2-p}\\
\vdots\\
(A_{0,s_{T-1}}+\Delta_{0,T-1})\psi_{T-1}\\
(A_{0,s_T}+\Delta_{0,T})\psi_T
\end{bmatrix},
\end{equation}
then the following Theorem, which is a trigger of pricing of the equity--linked life insurance products under the Bayesian MS--VAR process and which will be used in the rest of the paper holds.

\begin{thm}\label{thm02}
Let Bayesian MS--VAR($p$) process $y_t$ is given by equations \eqref{07001} or \eqref{07002}, for $t=1,\dots,T$, representation of random process $\theta_t$, which is $\mathcal{I}_{t-1}$ measurable is given by equation \eqref{07012}, and
\begin{equation}\label{07015}
\delta=\begin{bmatrix}
\delta_1\\ \delta_2
\end{bmatrix} ~~~\text{and}~~~
\Psi=\begin{bmatrix}
\Psi_{11} & 0\\
\Psi_{21} & \Psi_{22}
\end{bmatrix}
\end{equation}
be partitions, corresponding to random sub vectors $\bar{y}_t$ and $\bar{y}_t^c$ of the random vector $y$. Then the following probability laws hold:
\begin{eqnarray}
y~|~\mathcal{H}_0 &\sim& \mathcal{N}\Big(\Psi^{-1}\delta,\Psi^{-1}\Sigma(\Psi^{-1})'\Big), \label{07016}\\
\bar{y}_t~|~\mathcal{H}_0 &\sim& \mathcal{N}\Big(\Psi_{11}^{-1}\delta_1,\Psi_{11}^{-1}\bar{\Sigma}_t(\Psi_{11}^{-1})'\Big), \label{07017}\\
\bar{y}_t^c~|~\mathcal{H}_t &\sim& \mathcal{N}\Big(\Psi_{22}^{-1}\big(\delta_2-\Psi_{21}\bar{y}_t\big),\Psi_{22}^{-1}\bar{\Sigma}_t^c(\Psi_{22}^{-1})'\Big), \label{07018}\\ 
\tilde{q}_{x+k-1}&=&\frac{\exp{\{g_k\}}q_{x+k-1}}{p_{x+k-1}+\exp{\{g_k\}}q_{x+k-1}},~~~k=1,\dots,T\label{07019}
\end{eqnarray}
under the probability measure $\tilde{\mathbb{P}}$, where $C_{21}=-\Psi_{22}^{-1}\Psi_{21}\Psi_{11}^{-1}.$ Also, conditional on the initial information $\mathcal{F}_0$, a distribution of the random vector $\mathrm{vec}(s,\mathsf{P})$ is same for the risk--neutral probability measure $\mathbb{\tilde{P}}$ and the real probability measure $\mathbb{P}$ and for a conditional distribution of the random vector $\mathrm{vec}(\Pi_{\hat{s}},\Gamma_{\hat{s}})$, we have
\begin{equation}\label{07020}
\tilde{\mathbb{P}}\big[\mathrm{vec}(\Pi_{\hat{s}},\Gamma_{\hat{s}})\in B\big|s,\mathsf{P},\mathcal{F}_0\big]=\mathbb{P}\big[\mathrm{vec}(\Pi_{\hat{s}},\Gamma_{\hat{s}})\in B\big|\hat{s},\mathcal{F}_0\big],
\end{equation}
where $B\in\mathcal{B}(\mathbb{R}^d)$ with $d:=(np+k)nr_{\hat{s}}+(n_*p_*+k_*)n_*r_{\hat{s}}$ is a Borel set. Furthermore, random vectors $S:=S(s)=\mathrm{vec}(y,\Pi_{\hat{s}},\Gamma_{\hat{s}},s,\mathsf{P})$ and $I:=(I_1,\dots,I_T)'$ are independent under the risk--neutral probability measure $\mathbb{\tilde{P}}$.
\end{thm}

\begin{proof}
As the $\sigma$--fields $\mathcal{H}_T$ and $\mathcal{T}_T^x$ are independent under the real probability measure $\mathbb{P}$, the random vectors $S=\text{vec}(y,\Pi_{\hat{s}},\Gamma_{\hat{s}},s,\mathsf{P})$ and $I:=(I_1,\dots,I_T)^T$ are independent under the real probability measure $\mathbb{P}$. This means their joint distribution is represented by product measure, that is, $\mathbb{P}_{S,I}=\mathbb{P}_S\otimes \mathbb{P}_{I}$. Thus, for all Borel sets $B_1\in\mathcal{B}(\mathbb{R}^{Tn+\bar{d}})$ and $B_2\in\mathcal{B}(\mathbb{R}^T)$, by the Fubini Theorem we have
\begin{equation}\label{07021}
\mathbb{\tilde{P}}\big[(S,I)\in B_1\times B_2\big|\mathcal{F}_0\big]=\int_{B_1} L_T(z |\mathcal{F}_0)d\mathbb{P}_S[z |\mathcal{F}_0]\times \int_{B_2} K_T(i)d\mathbb{P}_{I}[i],
\end{equation}
where $\bar{d}:=(np+k)nr_{\hat{s}}+(n_*p_*+k_*)n_*r_{\hat{s}}+T+N(N+1)$ is a dimension of a random vector $\bar{S}:=\text{vec}(\Pi_{\hat{s}},\Gamma_{\hat{s}},s,\mathsf{P})$. Therefore, the random vector $S$ and random variable $I$ are independent under the probability measure $\mathbb{\tilde{P}}$. 

Let us take sets $B_1=\mathbb{R}^{Tn+\bar{d}}$ and $B_2=\{i\}$ in equation \eqref{07021}, where $i=(i_1,\dots,i_T)'\in \{0,1\}^T$ is a vector whose elements satisfy $0\leq i_1\leq \dots\leq i_T\leq 1$. Then it is clear that 
\begin{equation}\label{07022}
\mathbb{P}[I=i]=\begin{cases}
\mathbb{P}[T_x> T]={}_Tp_x & \text{if}~~~i_1=\dots=i_T=0,\\
\mathbb{P}[k-1< T_x\leq k]={}_{k-1}p_x-{}_kp_x & \text{if}~~~i_1=0,\dots,i_{k-1}=0,i_k=1,\dots,i_T=1,\\
0 & \text{if}~~~\text{otherwise}.
\end{cases}
\end{equation}
and
\begin{equation}\label{07023}
K_T(i)=\begin{cases}
\displaystyle \frac{\prod_{t=1}^T{}_{t-1}p_x}{\prod_{t=1}^T\big(\exp\{g_{t}\}{}_{t-1}p_x+(1-\exp\{g_{t}\}){}_tp_x\big)} & \text{if}~~~i_1=\dots=i_T=0,\\
\displaystyle \frac{\exp\{g_{k}\}\prod_{t=1}^k{}_{t-1}p_x}{\prod_{t=1}^k\big(\exp\{g_{t}\}{}_{t-1}p_x+(1-\exp\{g_{t}\}){}_tp_x\big)} & \text{if}~~~i_1=0,\dots,i_{k-1}=0,i_k=1,\dots,i_T=1,\\
0 & \text{if}~~~\text{otherwise}.
\end{cases}
\end{equation}
Since for each $\omega\in\{I=i\}$, a relation $K_T(\omega)=K_T(i)$ holds, it allows to conclude that $\mathbb{\tilde{P}}[I=i]=\int_{\{\omega:I=i\}} K_T(\omega)d\mathbb{P}_{T_x}[\omega]=K_T(i)\mathbb{P}[I=i]$. Therefore, one can obtain that 
\begin{equation}\label{07024}
{}_T\tilde{p}_x:=\mathbb{\tilde{P}}[T_x>T]=\frac{\prod_{t=1}^{T+1}{}_{t-1}p_x}{\prod_{t=1}^T\big(\exp\{g_{t}\}{}_{t-1}p_x+(1-\exp\{g_{t}\}){}_tp_x\big)}
\end{equation}
and for $k=1,\dots,T$,
\begin{equation}\label{07025}
{}_{k-1|}\tilde{q}_x:=\mathbb{\tilde{P}}[k-1<T_x\leq k]=\frac{\exp\{g_{k}\}({}_{k-1}p_x-{}_kp_x)\prod_{t=1}^k{}_{t-1}p_x}{\prod_{t=1}^k\big(\exp\{g_{t}\}{}_{t-1}p_x+(1-\exp\{g_{t}\}){}_tp_x\big)}
\end{equation}
and
\begin{equation}\label{07026}
{}_k\tilde{q}_x:=\mathbb{\tilde{P}}[T_x\leq k]=1-\frac{\prod_{t=1}^{k+1}{}_{t-1}p_x}{\prod_{t=1}^k\big(\exp\{g_{t}\}{}_{t-1}p_x+(1-\exp\{g_{t}\}){}_tp_x\big)}.
\end{equation}  
where $_T\tilde{p}_x$, $_{k-1|}\tilde{q}_x$ and $_k\tilde{q}_x$ represent probabilities that life-age-$x$ will attain age $x+T$, will survive $k-1$ years and die within the following 1 year, and will die within $k$ years, respectively under the risk--neutral probability measure $\mathbb{\tilde{P}}$. It follows from equation \eqref{07026} that
\begin{equation}\label{07027}
{}_{k-1}\tilde{p}_x-{}_k\tilde{p}_x={}_{k-1}\tilde{p}_x\frac{\exp\{g_k\}({}_{k-1}p_x-{}_kp_x)}{\exp\{g_k\}{}_{k-1}p_x+(1-\exp\{g_k\}){}_kp_x}.
\end{equation}
Consequently, we obtain equation \eqref{07019}:
\begin{equation}\label{07028}
\tilde{q}_{x+k-1}=\frac{{}_{k-1}\tilde{p}_x-{}_k\tilde{p}_x}{{}_{k-1}\tilde{p}_x}=\frac{\exp{\{g_k\}}q_{x+k-1}}{p_{x+k-1}+\exp{\{g_k\}}q_{x+k-1}}.
\end{equation}

For the rest of the proof, we follow \citeA{Battulga24a}. It is clear that conditional on information $\mathcal{H}_0$ a joint density function of the random vector $y=(y_1',\dots,y_T')'$ is given by
\begin{equation}\label{01147}
f(y|\mathcal{H}_0)=\frac{1}{(2\pi)^{nT/2}\prod_{t=1}^T|\Sigma_t|^{1/2}}\exp\bigg\{-\frac{1}{2}\sum_{t=1}^T\big(y_t-\Pi_{s_t} \mathsf{Y}_{t-1}\big)'\Sigma_t^{-1}\big(y_t-\Pi_{s_t} \mathsf{Y}_{t-1}\big)\bigg\}
\end{equation}
A joint density function of the random vector $S$ given $\mathcal{F}_0$ can be represented by
\begin{equation}\label{01148}
f\big(S |\mathcal{F}_0\big)=f(y |\mathcal{H}_0)\times f\big(\bar{S} |\mathcal{F}_0\big)
\end{equation}
under the real probability measure $\mathbb{P}$. Thus, conditional on initial information $\mathcal{F}_0$ a joint distribution of the random vector $S$ is given by
\begin{eqnarray}\label{01149}
&&\mathbb{\tilde{P}}\big[S\in B_1\big|\mathcal{F}_0\big]=\int_{B_1} L_T(y|\mathcal{F}_0)f\big(S |\mathcal{F}_0\big) dS\\
&&= \int_{B_1}c\exp\bigg\{-\frac{1}{2}\sum_{t=1}^T\big(y_t-\Pi_{s_t} \mathsf{Y}_{t-1}-\theta_t\big)'\Sigma_t^{-1}\big(y_t-\Pi_{s_t} \mathsf{Y}_{t-1}-\theta_t\big)\bigg\}\times f\big(\bar{S} |\mathcal{F}_0\big)dS\nonumber
\end{eqnarray}
under the probability measure $\tilde{\mathbb{P}}$, where the normalizing coefficient equals $c=\frac{1}{(2\pi)^{nT/2}\prod_{t=1}^T|\Sigma_t|^{1/2}}$ and $B_1\in\mathcal{B}(\mathbb{R}^{Tn+\bar{d}})$ is any Borel set. Therefore, one can conclude that conditional on $\mathcal{F}_0$ a joint distribution of the random vector $\bar{S}$ is same for probability measures $\mathbb{\tilde{P}}$ and $\mathbb{P}$, that is, for all Borel set $\bar{B}\in \mathcal{B}\big(\mathbb{R}^{\bar{d}}\big)$, 
\begin{equation}\label{01150}
\mathbb{\tilde{P}}\big[\bar{S}\in \bar{B}\big|\mathcal{F}_0\big]=\mathbb{P}\big[\bar{S}\in \bar{B}\big|\mathcal{F}_0\big].
\end{equation}
Since random vectors $\text{vec}(\Pi_{\hat{s}},\Gamma_{\hat{s}})$ and $\text{vec}(\mathsf{P})$ are independent given $s$ and $\mathcal{F}_0$ under the real probability measure $\mathbb{P}$, equation \eqref{01150} can be written by
\begin{equation}\label{01151}
\tilde{\mathbb{P}}\big[\text{vec}(\Pi_{\hat{s}},\Gamma_{\hat{s}})\in B\big|s,\mathsf{P},\mathcal{F}_0\big]\tilde{\mathbb{P}}\big[\text{vec}(s,\mathsf{P})\in D\big|\mathcal{F}_0\big]=\mathbb{P}\big[\text{vec}(\Pi_{\hat{s}},\Gamma_{\hat{s}})\in B\big|\hat{s},\mathcal{F}_0\big]\mathbb{P}\big[\text{vec}(s,\mathsf{P})\in D\big|\mathcal{F}_0\big],
\end{equation}
where $B\in\mathcal{B}(\mathbb{R}^d)$ and $D\in\mathcal{B}(\mathbb{R}^{T+N(N+1)})$ are Borel sets. Therefore, if we take $B=\mathbb{R}^d$ in above equation, then conditional on the initial information $\mathcal{F}_0$, distribution of the random vector $\text{vec}(s,\mathsf{P})$ is same for the both probability measures $\mathbb{\tilde{P}}$ and $\mathbb{P}$. Consequently, we have that
\begin{equation}\label{01152}
\tilde{\mathbb{P}}\big[\text{vec}(\Pi_{\hat{s}},\Gamma_{\hat{s}})\in B\big|s,\mathsf{P},\mathcal{F}_0\big]=\mathbb{P}\big[\text{vec}(\Pi_{\hat{s}},\Gamma_{\hat{s}})\in B\big|\hat{s},\mathcal{F}_0\big].
\end{equation}
Thus, for given regime--switching vector $s$, transition probability matrix $\mathsf{P}$, and initial information $\mathcal{F}_0$, a distribution of the random vector $\text{vec}(\Pi_{\hat{s}},\Gamma_{\hat{s}})$ under the risk--neutral probability measure $\mathbb{\tilde{P}}$ equals for given regime--switching vector $\hat{s}$ and initial information $\mathcal{F}_0$, a distribution of the random vector $\text{vec}(\Pi_{\hat{s}},\Gamma_{\hat{s}})$ under the real probability measure $\mathbb{P}$. Moreover, from equation \eqref{01149} we can conclude that conditional on information $\mathcal{H}_0$ a joint distribution of the random vector $y$ is given by
\begin{equation}\label{01153}
y~|~\mathcal{H}_0 \sim \mathcal{N}\Big(\Psi^{-1}\delta,\Psi^{-1}\Sigma(\Psi^{-1})'\Big)
\end{equation} 
under the risk--neutral probability measure $\tilde{\mathbb{P}}$. Thus equation \eqref{07016} holds. Thanks to the well--known formula of a partitioned matrix's inverse, we get that
\begin{equation}\label{•}
\Psi^{-1}=\begin{bmatrix}
\Psi_{11}^{-1} & 0\\
C_{21} & \Psi_{22}^{-1}
\end{bmatrix},
\end{equation}
where $C_{21}=-\Psi_{22}^{-1}\Psi_{21}\Psi_{11}^{-1}.$ Consequently, we obtain that
\begin{equation}\label{•}
\Psi^{-1}\delta=\begin{bmatrix}
\Psi_{11}^{-1}\delta_1\\
C_{21}\delta_1+\Psi_{22}^{-1}\delta_2
\end{bmatrix}
\end{equation}
and 
\begin{eqnarray}\label{•}
\Psi^{-1}\Sigma_s(\Psi^{-1})'=\begin{bmatrix}
\Psi_{11}^{-1}\bar{\Sigma}_t(\Psi_{11}^{-1})' & \Psi_{11}\bar{\Sigma}_t C_{21}'\\
C_{21}\bar{\Sigma}_t(\Psi_{11}^{-1})' & C_{21}\bar{\Sigma}_t C_{21}'+\Psi_{22}^{-1}\bar{\Sigma}_t^c(\Psi_{22}^{-1})'
\end{bmatrix}.
\end{eqnarray}
So equations \eqref{07017} holds. From the well--known formula of the conditional distribution of multivariate random vector, one can obtain equation \eqref{07018}. That completes the proof of the Theorem.
\end{proof}

\section{Log--normal System}

In this section, we briefly consider some results in \citeA{Battulga24a}. Because the idea of the domestic market can be used for domestic--foreign market, to simplify the calculation, here we will focus on the domestic market. We assume that financial variables, which are composed of a domestic log spot rate and domestic assets, and economic variables are together placed on Bayesian MS--VAR process $y_t$. To extract the financial variables from the process $y_t$, we introduce the following vectors and matrices: $e_i\in \mathbb{R}^n$ is an unit vector, whose $i$--th component is one and others are zero, and $J_z:=\big[I_{n_z}:0_{n_z\times n_x}\big]$ and $J_z:=\big[0_{n_x\times n_z}:I_{n_x}\big]$ matrices are can be used to extract processes $z_t$ and $x_t$ from the Bayesian MS--VAR$(p)$ process $y_t$, which is given by equations \eqref{07001} or \eqref{07002}.

Let $r_t$ be a domestic spot interest rate and $\tilde{r}_{t}:=\ln(1+r_t)$ be a log spot interest rate. Since the spot interest rate at time $t$ is known at time $(t-1)$, we can assume that the log spot rate is placed on the first component of the process $y_{t-1}$, i.e., $\tilde{r}_{t}=e_1^Ty_{t-1}$. Let $n_z\geq 1$ and $z_t:=J_zy_t$ be an $(n_z\times 1)$ vector at time $t$ that includes the domestic log spot rate. Since the first component of the process $z_t$ corresponds to the domestic log spot rate, we assume that other components of the process $z_t$ correspond to economic variables that affect domestic assets. So, the log spot rate is not constant and is explained by its own and other variables' lagged values in the VAR$(p)$ process $y_t$.

Henceforth, for a generic vector $o=(o_1,\dots,o_m)'\in\mathbb{R}^m$ and a function $f: \mathbb{R}\to \mathbb{R}$, we will use a vector notation: $f(o)=(f(o_1),\dots,f(o_m))'$. Let us suppose that $\tilde{x}_t:=\ln(x_t)=J_xy_t$ is an $(n_x\times 1)$ log price process of the domestic assets, where $x_t$ is an $(n_x\times 1)$ price process of the domestic assets. This means log prices of the domestic assets are placed on $(n_z+1)$--th to $(n=n_z+n_x)$--th components of the Bayesian MS--VAR$(p)$ process $y_t$. As a result, the domestic market is given by the following system:
\begin{equation}\label{07029}
\begin{cases}
z_t=\Pi_{1,s_t}\mathsf{Y}_{t-1}+\zeta_t\\
\tilde{x}_t=\Pi_{2,s_t}\mathsf{Y}_{t-1}+\eta_t\\
D_{t}=\exp\{-\tilde{r}_{1}-\tilde{r}_{2}-\dots-\tilde{r}_{t}\}=\frac{1}{\prod_{m=1}^t(1+r_m)}\\
\tilde{r}_{t}=e_1^Ty_{t-1}
\end{cases},~~~t=1,\dots,T,
\end{equation}
where $D_t$ is a discount process, $\zeta_t:=J_z\xi_t$ and $\eta_t:=J_x\xi_t$ are residual processes of the processes $z_t$ and $\tilde{x}_t$, respectively, $\Sigma_{11,t}:=J_z\Sigma_tJ_z'$, $\Sigma_{12,t}:=J_z\Sigma_tJ_x'$, $\Sigma_{21,t}:=J_x\Sigma_tJ_z'$, $\Sigma_{22,t}:=J_x\Sigma_tJ_x'$ be partitions, corresponding to the residual process $\xi_t=(\zeta_t',\eta_t')'$ of the covariance matrix $\Sigma_t$, and $\Pi_{1,s_t}:=J_z\Pi_{s_t}$ and $\Pi_{2,s_t}:=J_x\Pi_{s_t}$ are random coefficient matrices. For the system, $D_tx_t$ represents a discounted price process of the domestic assets. 

If we define a random vector $\hat{\theta}_{2,t}:=J_x(y_{t-1}-\Pi_{s_t}\mathsf{Y}_{t-1})+i_{n_x}e_1^Ty_{t-1}$, then it can be shown that
\begin{equation}\label{07030}
D_tx_t=\big(D_{t-1}x_{t-1}\big)\odot\exp\big(\eta_t-\hat{\theta}_{2,t}\big),
\end{equation}
where $\odot$ denotes the Hadamard product of two vectors. The random vector $\hat{\theta}_{2,t}$, which is measurable with respect to $\sigma$--field $\mathcal{I}_{t-1}$ is represented by
$$\hat{\theta}_{2,t}=\hat{\Delta}_{0,t}\psi_t+\hat{\Delta}_{1,t}y_{t-1}+\dots+\hat{\Delta}_{p,t}y_{t-p},$$
where $\hat{\Delta}_{0,t}:=-J_xA_{0,s_t}$, $\hat{\Delta}_{1,t}:=J_x\big(I_n-A_{1,s_t}\big)+i_{n_x}e_1^T$ and for $m=2,\dots,T$, $\hat{\Delta}_{m,t}:=-J_xA_{m,s_t}$. According to equation \eqref{07030} and the First Fundamental Theorem of asset pricing, as $D_{t-1}x_{t-1}$ is $\mathcal{H}_{t-1}$ measurable, in order to the discounted price process $D_tx_t$ is a martingale with respect to the filtration $\{\mathcal{H}_t\}_{t=0}^T$ and some risk--neutral probability measure $\tilde{\mathbb{P}}$, we must require that 
\begin{equation}\label{07031}
\tilde{\mathbb{E}}\big[\exp\big\{\eta_t-\hat{\theta}_{2,t}\big\}|\mathcal{H}_{t-1}\big]=i_{n_x},
\end{equation}
where $\mathbb{\tilde{E}}$ denotes an expectation under the risk--neutral probability measure $\tilde{\mathbb{P}}$. 
 
It is worth mentioning that condition \eqref{07031} corresponds only to the residual process $\eta_t$. Thus, we need to impose a condition on the residual processes $\zeta_t$ and death indicator process $I_t$ under the risk--neutral probability measure. Because for any admissible choices of the Girsanov kernel process $\theta_{1,t}$, corresponding to residual random process $\zeta_t$ and the vector $g$, condition \eqref{07031} holds, the market is incomplete. But net single premiums of the equity--linked life insurance products, which will be defined below are still consistent with the absence of arbitrage. For this reason, to price the equity--linked life insurance products, we will use the optimal Girsanov kernel process $\theta_t$ and the vector $g$, which minimize the variance of the state price density process at time $T$ and the relative entropy. According to \citeA{Battulga24a} and Theorem 1, the optimal Girsanov kernel process $\theta_t$ and the vector $g$ are obtained by
\begin{equation}\label{07032}
\theta_t=\Theta_t(\hat{\theta}_{2,t}-\alpha_{2,t}) ~~~\text{and}~~~g=0
\end{equation}
for $t=1,\dots,T$, where $\Theta_t:=\big[(\Sigma_{12,t}\Sigma_{22,t}^{-1})':I_{n_x}\big]'$, $\alpha_{2,t}:=\frac{1}{2}\mathcal{D}[\Sigma_{22,t}]$, and for generic square matrix $O$, $\mathcal{D}[O]$ denotes a vector, consisting of diagonal elements of the matrix $O$. The the risk--neutral probability measure $\tilde{\mathbb{P}}$, corresponding to equation \eqref{07032}, is called minimal martingale measure, see \citeA{Schweizer95}. 

Let us denote the first column of a generic matrix $O$ by $(O)_1$ and a matrix, which consists of other columns of the matrix $O$ by $(O)_1^c$. Then, the representation of the Girsanov kernel process of Theorem 1 is given by
\begin{equation}\label{07033}
\theta_t=\Delta_{0,t}\psi_t+\Delta_{1,t}y_{t-1}+\dots+\Delta_{p,t}y_{t-p},~~~t=1,\dots,T,
\end{equation}
where $\Delta_{0,t}:=[(\Delta_{0,t})_1:(\Delta_{0,t})_1^c]$ with $(\Delta_{0,t})_1=\Theta_t\big((\hat{\Delta}_{0,t})_1-\alpha_{2,t}\big)$ and $(\Delta_{0,t})_1^c=\Theta_t(\hat{\Delta}_{0,t})_1^c$, and for $m=1,\dots,p$, $\Delta_{m,t}:=\Theta_t\hat{\Delta}_{m,t}$. As a result, due to Theorem 1, conditional on $\mathcal{H}_t$, a distribution of the random vector $\bar{y}_t^c$ is given by
\begin{equation}\label{07034}
\bar{y}_t^c=(y_{t+1}',\dots,y_T')'~|~\mathcal{H}_t \sim \mathcal{N}\big(\mu_{2.1},\Sigma_{22.1}\big)
\end{equation}
under a risk--neutral probability measure (minimal martingale measure) $\tilde{\mathbb{P}}$, corresponding to the Girsanov kernel process \eqref{07033} and the vector $g=0$, where $\mu_{2.1}:=\Psi_{22}^{-1}\big(\delta_2-\Psi_{21}\bar{y}_t\big)$ and $\Sigma_{22.1}:=\Psi_{22}^{-1}\bar{\Sigma}_t^c(\Psi_{22}^{-1})'$ are mean vector and covariance matrix of the random vector $\bar{y}_t^c$ given $\mathcal{H}_t$, respectively. Note that the expectation $\mu_{2.1}$ and covariance $\Sigma_{22.1}$ are depend on the information $\mathcal{H}_t$.

According to \citeA{Geman95}, clever changes of probability measures lead to a significant reduction in the computational burden of derivative pricing. Therefore, we will consider the forward probability measure, originating from the risk--neutral probability measure $\mathbb{\tilde{P}}$. The forward measure is frequently used to price options, bonds, and interest rate derivatives. For this reason, we define the following domestic $(t,u)$--forward measure:
\begin{equation}\label{07035}
\mathbb{\hat{P}}_{t,u}\big[A\big|\mathcal{H}_t\big]:=\frac{1}{D_tB_{t,u}(\mathcal{H}_t)}\int_AD_u\mathbb{\tilde{P}}\big[\omega\big|\mathcal{H}_t\big],~~~\text{for all}~A\in \mathcal{H}_T
\end{equation}
where for given $\mathcal{H}_t$, $B_{t,u}(\mathcal{H}_t):=\frac{1}{D_t}\mathbb{\tilde{E}}[D_u|\mathcal{H}_t]$ is a price at time $t$ of a domestic zero--coupon bond paying 1 at time $u$.

Let us introduce a vector that deals with the domestic risk--free spot interest rate: for $t<u$, a vector $\gamma_{t,u}$ is defined by $\gamma_{t,u}':=\big[i_{u-t-1}'\otimes e_1':0_{1\times [(T-u+1)n]}\big]$, where $\otimes$ is the Kronecker product of two vectors. Then, we have that for $t<u$,
\begin{equation}\label{07036}
\sum_{m=t+1}^u\tilde{r}_{m}=\tilde{r}_{t+1}+\gamma_{t,u}'\bar{y}_t^c.
\end{equation}
According to equation \eqref{07034} and the completing square method, an exponent of a conditional expectation $\mathbb{\tilde{E}}\big[\frac{D_u}{D_t}\big|\mathcal{H}_t\big]$ can be represented by
\begin{eqnarray}\label{07037}
&&2\sum_{m=t+1}^u\tilde{r}_{m}+\big(\bar{y}_t^c-\mu_{2.1}\big)'\Sigma_{22.1}^{-1}\big(\bar{y}_t^c-\mu_{2.1}\big)\nonumber\\
&&=\Big(\bar{y}_t^c-\mu_{2.1}+\Sigma_{22.1}\gamma_{t,u}\Big)'\Sigma_{22.1}^{-1} \Big(\bar{y}_t^c-\mu_{2.1}+\Sigma_{22.1}\gamma_{t,u}\Big)\\
&&+2\Big(\tilde{r}_{t+1}+\gamma_{t,u}'\mu_{2.1}\Big)-\gamma_{t,u}'\Sigma_{22.1}\gamma_{t,u}\nonumber.
\end{eqnarray}
The second line of the above equation corresponds to price at time $t$ of the zero coupon bond. Consequently, for given $\mathcal{H}_t$, the price at time $t$ of the domestic zero--coupon bond with maturity $u$ is obtained as
\begin{equation}\label{07038}
B_{t,u}(\mathcal{H}_t)=\exp\bigg\{-\tilde{r}_{t+1}-\gamma_{t,u}'\mu_{2.1}+\frac{1}{2}\gamma_{t,u}'\Sigma_{22.1}\gamma_{t,u}\bigg\}.
\end{equation}
The first term of the exponent, which is given by equation \eqref{07037} corresponds a distribution of the random vector of the endogenous variables $\bar{y}_t^c$:
\begin{equation}\label{07040}
\bar{y}_t^c=(y_{t+1}',\dots,y_T')' ~|~\mathcal{H}_t \sim\mathcal{N}\Big(\hat{\mu}_{t,u},\Sigma_{22.1}\Big),
\end{equation}
under the $(t,u)$--forward measure $\mathbb{\hat{P}}_{t,u}$, where $\hat{\mu}_{t,u}:=\mu_{2.1}-\Sigma_{22.1}\gamma_{t,u}$ is an expectation of the random vector $\bar{y}_t^c$ under the forward probability measure. Note that because of the change of probability measure, the condition expectation of $\hat{\mu}_{t,u}$, corresponding to the $(t,u)$--forward probability measure $\hat{\mathbb{P}}_{t,u}$ is changed by $\Sigma_{22.1}\gamma_{t,u}$ from the conditional expectation $\mu_{2.1}$, corresponding to the risk--neutral probability measure $\tilde{\mathbb{P}}$, but the covariance matrix $\Sigma_{22.1}$ does not change for the two probability measures.

For a generic random vector $X$, we denote its joint density function by $\tilde{f}(X)$ under the risk--neutral probability measure $\mathbb{\tilde{P}}$ to differentiate the joint density function $f(X)$ under the real probability measure $\mathbb{P}$. To price the equity--linked life insurance products, we will use the following Lemmas.

\begin{lem}\label{lem01}
Conditional on $\mathcal{F}_t$, a joint density of $\big(\Pi_{\hat{s}},\Gamma_{\hat{s}},s,\mathsf{P}\big)$ is given by
\begin{equation}\label{07042}
\tilde{f}\big(\Pi_{\hat{s}},\Gamma_{\hat{s}},s,\mathsf{P}|\mathcal{F}_t\big)=\frac{\tilde{f}(\bar{y}_t|\Pi_{\alpha},\Gamma_{\alpha},\bar{s}_t,\mathcal{F}_0)f(\Pi_{\hat{s}},\Gamma_{\hat{s}}|\hat{s},\mathcal{F}_0)f(s,\mathsf{P}|\mathcal{F}_0)}{\displaystyle \sum_{\bar{s}_t}\bigg(\int_{\Pi_{\alpha},\Gamma_{\alpha}}\tilde{f}(\bar{y}_t|\Pi_{\alpha},\Gamma_{\alpha},\bar{s}_t,\mathcal{F}_0)f(\Pi_{\alpha},\Gamma_{\alpha}|\alpha,\mathcal{F}_0)d\Pi_\alpha d\Gamma_\alpha\bigg)f(\bar{s}_t|\mathcal{F}_0)}
\end{equation}
for $t=1,\dots,T$, where for $t=1,\dots,T$,
\begin{equation}\label{07043}
\tilde{f}(\bar{y}_t|\Pi_{\alpha},\Gamma_{\alpha},\bar{s}_t,\mathcal{F}_0)=\frac{1}{(2\pi)^{nt/2}|\Sigma_{11}|^{1/2}}\exp\Big\{-\frac{1}{2}\big(\bar{y}_t-\mu_1\big)'\Sigma_{11}^{-1}\big(\bar{y}_t-\mu_1\big)\Big\}
\end{equation}
with $\mu_1:=\Psi_{11}^{-1}\delta_1$ and $\Sigma_{11}:=\Psi_{11}^{-1}\bar{\Sigma}_t(\Psi_{11}^{-1})'$. In particular, we have that
\begin{equation}\label{ad001}
\tilde{f}\big(\Pi_{\hat{s}},\Gamma_{\hat{s}},s|\mathcal{F}_t\big)=\frac{\tilde{f}(\bar{y}_t|\Pi_{\alpha},\Gamma_{\alpha},\bar{s}_t,\mathcal{F}_0)f(\Pi_{\hat{s}},\Gamma_{\hat{s}}|\hat{s},\mathcal{F}_0)f(s|\mathcal{F}_0)}{\displaystyle \sum_{\bar{s}_t}\bigg(\int_{\Pi_{\alpha},\Gamma_{\alpha}}\tilde{f}(\bar{y}_t|\Pi_{\alpha},\Gamma_{\alpha},\bar{s}_t,\mathcal{F}_0)f(\Pi_{\alpha},\Gamma_{\alpha}|\alpha,\mathcal{F}_0)d\Pi_\alpha d\Gamma_\alpha\bigg)f(\bar{s}_t|\mathcal{F}_0)}
\end{equation}
for $t=1,\dots,T$.
\end{lem}

\begin{proof}
By the conditional probability formula, one gets that
\begin{equation*}\label{01162}
\tilde{f}(\bar{y}_t,\Pi_{\hat{s}},\Gamma_{\hat{s}},s,\mathsf{P}|\mathcal{F}_0)=\tilde{f}(\bar{y}_t|\Pi_{\hat{s}},\Gamma_{\hat{s}},s,\mathsf{P},\mathcal{F}_0)\tilde{f}(\Pi_{\hat{s}},\Gamma_{\hat{s}}|s,\mathsf{P},\mathcal{F}_0)\tilde{f}(s,\mathsf{P}|\mathcal{F}_0).
\end{equation*}
Due to Theorem 1, the first, second, and third terms of the right--hand side of the above equation equal $\tilde{f}(\bar{y}_t|\Pi_{\alpha},\Gamma_{\alpha},\bar{s}_t,\mathcal{F}_0)$, $f(\Pi_{\hat{s}},\Gamma_{\hat{s}}|\hat{s},\mathcal{F}_0)$, and $f(s,\mathsf{P}|\mathcal{F}_0)$, respectively. Consequently, by equation \eqref{08011}, we have that
\begin{equation}\label{01163}
\tilde{f}(\bar{y}_t,\Pi_{\hat{s}},\Gamma_{\hat{s}},s,\mathsf{P}|\mathcal{F}_0)=\tilde{f}(\bar{y}_t|\Pi_{\alpha},\Gamma_{\alpha},\bar{s}_t,\mathcal{F}_0)f(\Pi_{\alpha},\Gamma_{\alpha}|\alpha,\mathcal{F}_0)f_*(\Pi_{\delta},\Gamma_{\delta}|\delta,\mathcal{F}_0)f(s,\mathsf{P}|\mathcal{F}_0).
\end{equation}
If we take integral from the above equation with respect to $(\Pi_{\hat{s}},\Gamma_{\hat{s}},s,\mathsf{P})$, then we find conditional density of the random vector $\bar{y}_t$ under the risk--neutral probability $\mathbb{\tilde{P}}$, namely,
\begin{eqnarray}\label{01166}
\tilde{f}(\bar{y}_t|\mathcal{F}_0)=\sum_{\bar{s}_t}\bigg(\int_{\Pi_{\alpha},\Gamma_{\alpha}}\tilde{f}(\bar{y}_t|\Pi_{\alpha},\Gamma_{\alpha},\bar{s}_t,\mathcal{F}_0)f(\Pi_{\alpha},\Gamma_{\alpha}|\alpha,\mathcal{F}_0)d\Pi_\alpha d\Gamma_\alpha\bigg)f(\bar{s}_t|\mathcal{F}_0).
\end{eqnarray}
Dividing equation \eqref{01163} by equation \eqref{01166}, one obtains equation \eqref{07042}. If we integrate equation \eqref{07042} by $\mathsf{P}$, then we get equation \eqref{ad001}.
\end{proof}

\begin{lem}\label{lem02}
Let $X\in \mathbb{R}^n$ be a random vector with a distribution $X\sim\mathcal{N}(\mu,\Sigma)$, then for each deterministic matrix $A\in\mathbb{R}^{m\times n}$ ($m\leq n$), vectors $\alpha'\in \mathbb{R}^n$ and $b,c\in\mathbb{R}^m$, and number $\beta\in\mathbb{R}$, it holds
\begin{eqnarray}\label{07044}
\mathbb{E}\Big[\exp\big\{\alpha X+\beta\big\}1_{\{AX+b> c\}}\Big]=\exp\bigg\{\alpha\mu+\beta+\frac{1}{2}\alpha\Sigma\alpha'\bigg\}\mathbb{P}\Big[AX+b> c-A\Sigma\alpha'\Big].
\end{eqnarray}
\end{lem}

\begin{proof}
According to the completing the square method, we have
\begin{eqnarray}\label{04124}
&&\alpha x+\beta-\frac{1}{2}(x-\mu)'\Sigma^{-1}(x-\mu)\nonumber\\
&&=\alpha \mu+\beta+\frac{1}{2}\alpha\Sigma\alpha'-\frac{1}{2}\big(x-\mu-\Sigma\alpha'\big)'\Sigma^{-1}\big(x-\mu-\Sigma\alpha'\big).
\end{eqnarray}
Consequently, one finds that
\begin{equation}\label{04125}
\mathbb{E}\Big[\exp\big\{\alpha X+\beta\big\}1_{\{AX+b> c\}}\Big]=\exp\bigg\{\alpha \mu+\beta+\frac{1}{2}\alpha\Sigma\alpha'\bigg\}\mathbb{E}\Big[1_{\{AY+b> c\}}\Big],
\end{equation}
where $Y\sim\mathcal{N}(\mu+\Sigma \alpha',\Sigma)$. Since $Y\overset{d}{=}X+\Sigma\alpha'$, $\mathbb{E}\big[1_{\{AY+b> c\}}\big]=\mathbb{P}\big[AY+b> c\big]$, and $\mathbb{E}\big[\exp\big\{\alpha X+\beta\big\}\big]=\exp\big\{\alpha\mu+\beta+\frac{1}{2}\alpha\Sigma\alpha'\big\}$ we obtain equation \eqref{07044}, where $d$ denotes equal distribution.
\end{proof}

\section{Equity--Linked Life Insurance Products}

Now we consider the pricing of some equity--linked life insurance products on a maximum of several assets, which are connected to call and put rainbow options using the risk--neutral probability measure. Here we will price segregated fund contracts with guarantee, see \citeA{Hardy01} and unit--linked life insurance products with guarantee. We assume that operational expenses, which are deducted from a fund, and withdrawals are omitted from the life insurance products. Unit--linked life insurance contracts are very popular in many markets. For the contract, a return obtained by the policyholder on its savings is linked to a specific financial index. Typically, the policyholder receives a maximum of stock price and some guarantee, but various types of guarantees can be added to the pure unit--linked contract. For early theoretical analysis of unit--linked contracts, see \citeA{Aase94}, \citeA{Brennan76}, and \citeA{Moller98}. A common life insurance product in practice is endowment insurance and one can obtain that combinations of term life insurance and pure endowment insurance lead to numerous endowment insurances, see \citeA{Aase94}. Because almost all typical life insurance products can be expressed by in terms of term life insurance and pure endowment insurance, it is sufficient to consider only the term life insurance and pure endowment insurance.

Let $G_t$ be the amount of the guarantee at time $t$ and $n_x$ be the number of domestic assets. We define a maximum of weighted prices at time $t$ of the domestic assets by
\begin{equation}\label{07045}
M_t:=\max\{w_{1,t}x_{1,t},\dots,w_{n_x,t}x_{n_x,t}\}, ~~~~t=1,\dots,T,
\end{equation}
where $w_{i,t}$ is a weight and $x_{i,t}$ is the price, respectively, at time $t$ of $i$-th asset. One choice of the weights corresponds to the reciprocal of the prices at time 0 of the assets. In this case, $w_{i,t}x_{i,t}=x_{i,t}/x_{i,0}$ represents total return at time $t$ of $i$--th asset. Let $w_t:=(w_{1,t},\dots,w_{n_x,t})^T$ be a weight vector at time $t$ of the assets.

A $T$--year pure endowment insurance provides payment of the sum insured at the end of the $T$ years only if the insured is alive at the end of $T$ years from the time of policy issue. For the pure endowment insurance, we assume that the sum insured is of the form $f(M_T)$ for some Borel measurable function $f$, where $M_T$ is given by equation \eqref{07045}. Forms of the function $f$ depend on insurance contracts and choices of $f$ give us different types of life insurance products. For example, for $G_t>0$, $f(M_t)=1$, $f(M_t)=M_t$, $f(M_t)=\max\{M_t,G_t\}=[M_t-G_t]^++G_t$ and $f(M_t)=[M_t-G_t]^+$ correspond to a simple life insurance, pure unit--linked, unit linked with guarantee, and segregated fund contract with guarantee, respectively, see \citeA{Aase94}, \citeA{Bowers97}, \citeA{Hardy01} and \citeA{Moller98}. As a result, discounted contingent claim of the $T$--year pure endowment insurance can be represented by following equation
\begin{equation}\label{07046}
\overline{H}_T=D_Tf(M_T)1_{\{T_x>T\}}.
\end{equation}
Since $\sigma$--fields $\mathcal{H}_T$ and $\mathcal{T}_T^x$ are independent under the risk-neutral probability measure $\mathbb{\tilde{P}}$, one can obtain that conditional on the information $\mathcal{H}_t$, value at time $t$ of a contingent claim $H_T=f(M_T)1_{\{T_x>T\}}$ is given by
\begin{equation}\label{07047}
V_t(\mathcal{H}_t^x)=\frac{1}{D_t}\mathbb{\tilde{E}}[\overline{H}_T|\mathcal{H}_t^x]=1_{\{T_x>t\}}\frac{1}{D_t}\mathbb{\tilde{E}}[D_Tf(M_T)|\mathcal{H}_t]{}_{T-t}p_{x+t}.
\end{equation}

A $T$--year term life insurance is an insurance, which provides a payment of the sum insured only if death occurs in $T$ years. In contrast to pure endowment insurance, the term life insurance's sum insured depends on time $t$, that is, its sum insured is form $f(M_t)$ because random death occurs any time in $T$ years. Therefore, discounted contingent claim of the $T$--term life insurance is given by
\begin{equation}\label{07048}
\overline{H}_T=D_{K_x+1}f(M_{K_x+1})1_{\{K_x+1\leq T\}}=\sum_{k=0}^{T-1}D_{k+1}f(M_{k+1})1_{\{K_x=k\}},
\end{equation}
where $K_x:=[T_x]$ is the curtate future lifetime random variable of life--aged--$x$. For a contingent claim of the term life insurance, providing a benefit at the end of the year of death, it follows from the fact that $\mathcal{H}_T$ and $\mathcal{T}_T^x$ are independent that for the term insurance, its value process at time $t$ is
\begin{equation}\label{07049}
V_t(\mathcal{H}_t^x)=\frac{1}{D_t}\mathbb{\tilde{E}}[\overline{H}_T|\mathcal{H}_t^x]=1_{\{T_x>t\}}\sum_{k=t}^{T-1}\frac{1}{D_t}\mathbb{\tilde{E}}[D_{k+1}f(M_{k+1})|\mathcal{H}_t]{}_{k-t}p_{x+t}q_{x+k}.
\end{equation}

For the $T$--year term life insurance and $T$--year pure endowment insurance both of which correspond to the segregated fund contract, observe that the sum insured form is $f(M_k)=\big[G_k-M_k\big]^+$ for $k=1,\dots,T$. On the other hand, the form of the sum insured of the unit--linked life insurances is $f(M_k)=\big[M_k-G_k\big]^++G_k$ for $k=1,\dots,T$. Considering the structure of the sum insureds of the segregated funds and unit--linked life insurances, one can deduce that to price the life insurance products, it will be sufficient to consider European call and put options with strike price $G_k$ and maturity $k$ for $k=t+1,\dots,T$. 

To obtain a net single premium at time $t$ of the equity--linked life insurance products, we need a distribution of the log price process $\tilde{x}_k$. To keep notations simple, let $J_{k|t}^x:=[0:I_{n_x}:0]$ be an $(n_x\times [n(T-t)])$ matrix, whose $(k-t)$ block matrix equals $I_{n_x}$ and others are zero and the matrix is used to extract the log price vector $\tilde{x}_k$ from the random vector $\bar{y}_t^c$. Then, by equation \eqref{07040}, conditional on information $\mathcal{H}_t$, a distribution of the log price vector $\tilde{x}_k$ is given by
\begin{equation}\label{07050}
\tilde{x}_k~|~\mathcal{H}_t\sim \mathcal{H}\Big(\hat{\mu}_{k|t,u},\Sigma_{k|t}\Big),
\end{equation}
under the $(t,u)$--forward probability measure $\mathbb{\hat{P}}_{t,u}$, where $\hat{\mu}_{k|t,u}=J_{k|t}^x\hat{\mu}_{t,u}$ and $\Sigma_{k|t}=J_{k|t}^x\Sigma_{22.1}(J_{k|t}^x)'$ are expectation and covariance matrix of the log price vector $\tilde{x}_k$ for given $\mathcal{H}_t$, respectively.

Let us consider a call option at time $t$ with a payoff $(M_k-G_k)^+$ for $k=t+1,\dots,T$. The call option is the European call option on a maximum of asset prices with strike price $G_k$ and maturity $k$ and it is called the rainbow option. Then its discounted contingent claim can be represented by
\begin{equation}\label{07051}
D_k(M_k-G_k)^+=\sum_{i=1}^{n_x}D_k(Z_{i,k}-G_k)1_{E_{i,k}^c}
\end{equation}
where the event $E_{i,k}^c=\{Z_{i,k}\geq Z_{1,k},\dots,Z_{i,k}\geq Z_{i-1,k},Z_{i,k}\geq Z_{i+1,k},\dots,Z_{i,k}\geq Z_{n_x,k},Z_{i,k}\geq G_k\}$ represents $Z_{i,k}$ is a maximum of weighted asset prices at time $k$ and the call option expires in the money and $Z_{i,k}:=w_{i,k}x_{i,k}$ is the weighted asset price. In terms of the log price vector at time $k$ of the assets, $\tilde{x}_k$, the event $E_{i,k}^c$ can be written by 
\begin{equation}\label{07052}
E_{i,k}^c=\big\{L_i^c \tilde{x}_k\geq b_{i,k}^c\big\}
\end{equation}
where $b_{i,k}^c:=\big(\ln(w_{1,k}/w_{i,k}),\dots,\ln(w_{i-1,k}/w_{i,k}),\ln(w_{i+1,k}/w_{i,k}),\dots,\ln(w_{n_x,k}/w_{i,k}),\ln(G_k/w_{i,k})\big)'$ and 
\begin{equation}\label{07053}
L_i^c:=\begin{bmatrix}
-I_{i-1} & i_{i-1} & 0\\
0 & i_{n_x-i} & -I_{n_x-i}\\
0 & 1 & 0
\end{bmatrix}.
\end{equation}
Then, it follows from the $(t,k)$--forward measure, equations \eqref{07051}--\eqref{07053}, and Lemma \ref{lem02} that conditional on $\mathcal{H}_t$, a price of the call option is obtained by
\begin{eqnarray}\label{07054}
C_{k|t}(\mathcal{H}_t)&=&\frac{1}{D_t}\mathbb{\tilde{E}}\Big[D_k\big(M_k-G_k\big)^+\Big|\mathcal{H}_t\Big]\nonumber\\
&=&B_{t,k}(\mathcal{H}_t)\sum_{i=1}^{n_x}\bigg(w_{i,k}\exp\bigg\{e_i'\hat{\mu}_{k|t,k}+\frac{1}{2}e_i'\Sigma_{k|t}e_i\bigg\}\nonumber\\
&\times & \bar{\Phi}_{n_x}\Big(\big[L_i^c\Sigma_{k|t}(L_i^c)'\big]^{-1/2}\big(b_{i,k}^c-L_i^c\Sigma_{k|t}e_i-L_i^c\hat{\mu}_{k|t,k}\big)\Big)\\
&- & G_k\bar{\Phi}_{n_x}\Big(\big[L_i^c\Sigma_{k|t}(L_i^c)'\big]^{-1/2}\big(b_{i,k}^c-L_i^c\hat{\mu}_{k|t,k}\big)\Big)\bigg),\nonumber
\end{eqnarray}
where for a random vector $Z\sim \mathcal{N}(0,I_n)$, $\bar{\Phi}_n(z):=\mathbb{P}[Z\geq z]=\prod_{i=1}^n(1-\Phi(z_i))$ is a survival probability and $\Phi(z_i):=\int_{-\infty}^{z_1}\frac{1}{2\pi}e^{-t^2/2}dt$ is the cumulative distribution function of a standard normal random variable.
As a result, due to Lemma \ref{lem01} and the tower property of conditional expectation, the price at time $t$ of the call option on a maximum of the prices at time $k$ of the assets with maturity $k$ and strike price $G_k$ is obtained by 
\begin{eqnarray}\label{07056}
C_{k|t}=\frac{1}{D_t}\tilde{\mathbb{E}}\big[D_k\big(M_k-G_k\big)^+\big|\mathcal{F}_t\big]=\sum_{s}\int_{\Pi_{\hat{s}},\Gamma_{\hat{s}}}C_{k|t}(\mathcal{H}_t)\tilde{f}(\Pi_{\hat{s}},\Gamma_{\hat{s}},s|\mathcal{F}_t)d\Pi_{\hat{s}} d\Gamma_{\hat{s}}.
\end{eqnarray}

Similarly to the call option, it can be shown that for $k=t+1,\dots,T$, conditional on information $\mathcal{H}_t$, price at time $t$ of a put option on maximum is given by
\begin{eqnarray}\label{07057}
P_{k|t}(\mathcal{H}_t)&=&\frac{1}{D_t}\mathbb{\tilde{E}}\Big[D_k\big(G_k-M_k\big)^+\Big|\mathcal{H}_t\Big]\nonumber\\
&=&B_{t,k}(\mathcal{H}_t)\sum_{i=1}^{n_x}\bigg(G_k\bar{\Phi}_{n_x}\Big(\big[L_i^p\Sigma_{k|t}(L_i^p)'\big]^{-1/2}\big(b_{i,k}^p-L_i^c\hat{\mu}_{k|t,k}\big)\Big)\nonumber\\
&-&w_{i,k}\exp\bigg\{e_i'\hat{\mu}_{k|t,k}+\frac{1}{2}e_i'\Sigma_{k|t}e_i\bigg\}\\
&\times & \bar{\Phi}_{n_x}\Big(\big[L_i^p\Sigma_{k|t}(L_i^p)'\big]^{-1/2}\big(b_{i,k}^p-L_i^p\Sigma_{k|t}e_i-L_i^p\hat{\mu}_{k|t,k}\big)\Big)\bigg).\nonumber
\end{eqnarray}
where $b_{i,k}^p:=\big(\ln(w_{1,k}/w_{i,k}),\dots,\ln(w_{i-1,k}/w_{i,k}),\ln(w_{i+1,k}/w_{i,k}),\dots,\ln(w_{n_x,k}/w_{i,k}),\ln(w_{i,k}/G_k)\big)'$ and 
\begin{equation}\label{07058}
L_i^p:=\begin{bmatrix}
-I_{i-1} & i_{i-1} & 0\\
0 & i_{n_x-i} & -I_{n_x-i}\\
0 & -1 & 0
\end{bmatrix}.
\end{equation}
Therefore, due to Lemmas \ref{lem01} and the tower property of conditional expectation, the price at time $t$ of the put option on a maximum of the prices at time $k$ of the assets with maturity $k$ and strike price $G_k$ is obtained by
\begin{eqnarray}\label{07059}
P_{k|t}=\frac{1}{D_t}\tilde{\mathbb{E}}\big[D_k\big(G_k-M_k\big)^+\big|\mathcal{F}_t\big]=\sum_s\int_{\Pi_{\hat{s}},\Gamma_{\hat{s}}}P_{k|t}(\mathcal{H}_t)\tilde{f}(\Pi_{\hat{s}},\Gamma_{\hat{s}},s|\mathcal{F}_t)d\Pi_{\hat{s}} d\Gamma_{\hat{s}}.
\end{eqnarray}

As a result, it follows from equations \eqref{07056} and \eqref{07059} that single premiums of the $T$--year equity--linked life insurance products without withdrawal and operational expenses, providing a benefit at the end of the year of death (term life insurance) and the end of the $T$ years (pure endowment insurance), are given by
\begin{itemize}
\item[1.] for the $T$--year guaranteed term life insurance, corresponding to segregated fund contract on a maximum of weighted prices of the domestic assets, it holds
\begin{equation}\label{07060}
\lcterm{S}{x+t}{T-t}=1_{\{T_x>t\}}\sum_{k=t}^{T-1}P_{k+1|t}{}_{k-t}p_{x+t}q_{x+k};
\end{equation}
\item[2.] for the $T$--year guaranteed pure endowment insurance, corresponding to segregated fund contract on a maximum of weighted prices of the domestic assets, it holds
\begin{equation}\label{07061}
\lcend{S}{x+t}{T-t}=1_{\{T_x>t\}}P_{T|t}{}_{T-t}p_{x+t};
\end{equation}
\item[3.] for the $T$--year guaranteed unit--linked term life insurance on a maximum of weighted prices of the domestic assets, it holds
\begin{equation}\label{07062}
\lcterm{U}{x+t}{T-t}=1_{\{T_x>t\}}\sum_{k=t}^{T-1}\Big(C_{k+1|t}+G_{k+1}B_{t,k+1}\Big){}_{k-t}p_{x+t}q_{x+k};
\end{equation}
\item[4.] for the $T$--year guaranteed unit--linked pure endowment insurance on a maximum of weighted prices of the domestic assets, it holds
\begin{equation}\label{07063}
\lcend{U}{x+t}{T-t}=1_{\{T_x>t\}}\Big(C_{T|t}+G_TB_{t,T}\Big){}_{T-t}p_{x+t},
\end{equation}
\end{itemize}
where for $k=t+1,\dots,T$, $B_{t,k}(\mathcal{F}_t)$ is zero--coupon bond price at time $t$, which expires at time $k$ and is obtained by the following equation 
\begin{equation}\label{07064}
B_{t,k}:=\sum_{s}\int_{\Pi_{\hat{s}},\Gamma_{\hat{s}}}B_{t,k}(\mathcal{H}_t)\tilde{f}(\Pi_{\hat{s}},\Gamma_{\hat{s}},s|\mathcal{F}_t)d\Pi_{\hat{s}} d\Gamma_{\hat{s}}.
\end{equation}

\section{Locally Risk--Minimizing Strategy}

The concept of locally risk--minimizing strategy goes back to \citeA{Follmer86}. According to \citeA{Follmer86}, if a discounted cumulative cost process is a martingale, then a portfolio plan is called mean--self--financing. In a discrete--time case, \citeA{Follmer89} developed a locally risk--minimizing strategy and obtained a recurrence formula for optimal strategy. According to \citeA{Schal94} (see also \citeA{Follmer04}), under a martingale probability measure the locally risk--minimizing strategy and remaining conditional risk--minimizing strategy are the same. Therefore, in this section, we will consider locally risk--minimizing strategies, which correspond to the equity--linked life insurance products given in Section 4. In the insurance industry, for continuous--time unit--linked term life and pure endowment insurances with guarantee, locally risk--minimizing strategies are obtained by \citeA{Moller98}.

To simplify notations we define: for $t=1,\dots,T$, $\overline{X}_t:=(\overline{X}_{1,t},\dots,\overline{X}_{n,t})'$ is a discounted price process at time $t$ and $\Delta \overline{X}_t:=\overline{X}_t-\overline{X}_{t-1}$ is a difference process at time $t$ of the discounted price process, where $\overline{X}_{i,t}:=D_tx_{i,t}$ is a discounted price process at time $t$ of $i$--th stock. Note that $\Delta \overline{X}_t$ is a martingale difference with respect to the filtration $\{\mathcal{H}_t\}_{t=0}^T$ and the risk--neutral probability measure $\mathbb{\tilde{P}}$. Also, let $h_t$ be a proper number of shares at time $t$ and $h_t^0$ be a proper amount of cash (risk--free bond) at time $t$, which are required to successfully hedge a generic contingent claim $H_T$, and $\overline{H}_T$ be a discounted contingent claim, where we assume that the discounted contingent claim $\overline{H}_T$ is square--integrable under the risk--neutral probability measure. Then, following the idea in \citeA{Follmer04} and \citeA{Follmer89}, one can obtain that for a filtration $\{\mathcal{F}_t^x\}_{t=0}^T$ and the generic discounted contingent claim $\overline{H}_T$, under the risk--neutral probability measure $\mathbb{\tilde{P}}$, the locally risk--minimizing strategy ($h^0, h$) is given by the following equations:
\begin{equation}\label{07065}
h_{t+1}=\Omega_{t+1}^{-1}\Lambda_{t+1}~~~\text{and}~~~h_{t+1}^0=V_{t+1}-h_{t+1}'x_{t+1}
\end{equation}
for $t=0,\dots,T-1$ and $h_0^0=V_0-h_1'x_1$, where $\Omega_{t+1}:=\tilde{\mathbb{E}}\big[\Delta\overline{X}_{t+1}\Delta\overline{X}_{t+1}'\big|\mathcal{F}_{t}^x\big]$ is an $(n_x\times n_x)$ random matrix, $\Lambda_{t+1}:=\widetilde{\text{Cov}}\big[\Delta\overline{X}_{t+1},\overline{H}_T\big|\mathcal{F}_{t}^x\big]$ is an $(n_x\times n_x)$ random matrix, and $V_{t+1}:=\frac{1}{D_{t+1}}\mathbb{\tilde{E}}[\overline{H}_T|\mathcal{F}_{t+1}^x]$ is a value process of the contingent claim. Note that $h_t$ is a predictable process, which means its value is known at time $t-1$, while for the process $h_t^0$, its value is only known at time $t$, and if the contingent claim $H_T$ is generated by stock price process $P_t$ for $t=1,\dots,T$, then the process $h_t^0$ becomes predictable, see \citeA{Follmer89}. Note that for $t=1,\dots,T$, since $\sigma$--fields $\mathcal{H}_T$ and $\mathcal{T}_T^x$ are independent, if $X$ is any random variable, which is independent of $\sigma$--field $\mathcal{T}_t^x$ and integrable with respect to the risk--neutral probability measure, then it holds
\begin{equation}\label{07066}
\mathbb{\tilde{E}}[X|\mathcal{H}_t^x]=\mathbb{\tilde{E}}[X|\mathcal{H}_t].
\end{equation}

Because under the risk--neutral probability measure $\mathbb{\tilde{P}}$ the difference process $\Delta\overline{X}_t$ of discounted price process $\overline{X}_t$ is independent of $\sigma$--field $\mathcal{T}_t^x$ and is a martingale difference with respect to filtration $\{\mathcal{H}_t\}_{t=0}^T$ and the risk--neutral probability measure $\tilde{\mathbb{P}}$, it follows that
\begin{equation}\label{07067}
\Omega_{t+1}(\mathcal{H}_t):=\tilde{\mathbb{E}}\big[\Delta\overline{X}_{t+1}\Delta\overline{X}_{t+1}'\big|\mathcal{H}_{t}^x\big]=\mathbb{\tilde{E}}\big[\overline{X}_{t+1}\overline{X}_{t+1}'\big|\mathcal{H}_{t}\big]-\overline{X}_{t}\overline{X}_{t}'
\end{equation}
and
\begin{equation}\label{07068}
\Lambda_{t+1}(\mathcal{H}_t^x):=\widetilde{\text{Cov}}\big[\Delta\overline{X}_{t+1},\overline{H}_T\big|\mathcal{H}_{t}^x\big]=\mathbb{\tilde{E}}\big[\overline{H}_T\overline{X}_{t+1}\big|\mathcal{H}_t^x\big]-\overline{V}_{t}(\mathcal{H}_t^x)\overline{X}_{t}
\end{equation}
for $t=0,\dots,T-1$. 

Let us consider equation \eqref{07067}. Since the discounted price process $\overline{X}_t$ is a martingale with respect to the filtration $\{\mathcal{H}_t\}_{t=0}^T$ and the risk--neutral probability measure $\tilde{\mathbb{P}}$, by the well--known covariance formula of a multivariate log--normal random vector, we have that 
\begin{equation}\label{07071}
\Omega_{t+1}(\mathcal{H}_t)=\exp\big\{\Sigma_{t+1|t}\big\}\odot\overline{X}_{t}\overline{X}_{t}'.
\end{equation}
Consequently, it follows from the tower property of conditional expectation and Lemmas \ref{lem01} that the matrix $\Omega_{t+1}$ is obtained by  
\begin{eqnarray}\label{07072}
\Omega_{t+1}=\sum_{s}\int_{\Pi_{\hat{s}},\Gamma_{\hat{s}}}\Omega_{t+1}(\mathcal{H}_t)\tilde{f}(\Pi_{\hat{s}},\Gamma_{\hat{s}},s|\mathcal{F}_t)d\Pi_{\hat{s}} d\Gamma_{\hat{s}}.
\end{eqnarray}

Since the $\sigma$--fields $\mathcal{H}_T$ and $\mathcal{T}_T^x$ are independent, the conditional covariance $\Lambda_{t+1}(\mathcal{H}_t^x)$ equals
\begin{itemize}
\item[($i$)] for equity--linked pure endowment insurance, 
\begin{equation}\label{07073}
\Lambda_{t+1}(\mathcal{H}_t^x)=1_{\{T_x>t\}}\mathbb{\tilde{E}}\big[D_Tf(P_T)\overline{X}_{t+1}\big|\mathcal{H}_t\big]{}_{T-t}p_{x+t}-\overline{V}_{t}(\mathcal{H}_t^x)\overline{X}_{t}
\end{equation}
\item[($ii$)] for equity--linked term life insurance, 
\begin{equation}\label{07074}
\Lambda_{t+1}(\mathcal{H}_t^x)=1_{\{T_x>t\}}\sum_{k=t}^{T-1}\mathbb{\tilde{E}}\big[D_{k+1}f(P_{k+1})\overline{X}_{t+1}\big|\mathcal{H}_t\big]{}_{k-t}p_{x+t}q_{x+k}-\overline{V}_{t}(\mathcal{H}_t^x)\overline{X}_{t},
\end{equation}
\end{itemize}
where for the segregated fund contracts the sum insureds are $f(P_k)=[G_k-M_k]^+$ and for the unit--linked life insurances the sum insureds are $f(P_k)=[M_k-G_k]^++G_k$ for $k=t+1,\dots,T$. Note that integrals with respect to the vector $\bar{S}=\text{vec}(\Pi_{\hat{s}},\Gamma_{\hat{s}},s,\mathsf{P})$ from the discounted value processes $\overline{V}_{t}(\mathcal{H}_t^x)$ are obtained from equations \eqref{07060}--\eqref{07063}.

Let us consider the term $\mathbb{\tilde{E}}\big[D_kf(P_k)\overline{X}_{t+1}\big|\mathcal{H}_t\big]$ in equations \eqref{07073} and \eqref{07074}. For the segregate funds, according to $(t,k)$--forward probability measure, for $k=t+1,\dots,T$, the term is represented by
\begin{eqnarray}\label{07075}
&&S_{k|t}(\mathcal{H}_t):=\mathbb{\tilde{E}}\big[D_k(G_k-M_k)^+\overline{X}_{t+1}\big|\mathcal{H}_t\big]=D_tD_{t+1}B_{t,k}(\mathcal{H}_t)\\
&&\times\sum_{i=1}^{n_x}\bigg(G_k\mathbb{\hat{E}}_{t,k}\Big[\exp{\{\tilde{x}_{t+1}\}}1_{E_{i,k}^p}\Big|\mathcal{H}_t\Big]-w_{i,k}\mathbb{\hat{E}}_{t,k}\Big[\exp{\{\tilde{x}_{t+1}+\tilde{x}_{i,k}i_{n_x}\}}1_{E_{i,k}^p}\Big|\mathcal{H}_t\Big]\bigg),\nonumber
\end{eqnarray}
where $\hat{\mathbb{E}}_{t,k}$ is an expectation under the $(t,k)$--forward probability measure $\hat{\mathbb{P}}_{t,k}$, According to Lemma \ref{lem02}, for $j=1,\dots,n_x$, $j$--th component of the last expectation of the above equation is given by the following equation
\begin{eqnarray}\label{07077}
&&\mathbb{\hat{E}}_{t,k}\Big[\exp{\{\tilde{x}_{j,t+1}+\tilde{x}_{i,k}\}}1_{E_{i,k}^p}\Big|\mathcal{H}_t\Big]\nonumber\\
&&=\exp\bigg\{e_j'\hat{\mu}_{t+1|t,k}+e_i'\hat{\mu}_{k|t,k}+\frac{1}{2}e_j'\Sigma_{t+1|t}e_j+\frac{1}{2}e_i'\Sigma_{k|t}e_i+e_j'J_{t+1|t}^x\Sigma_{22.1}(J_{k|t}^x)'e_i\bigg\}\\
&&\times\bar{\Phi}_{n_x}\Big(\big[L_i^p\Sigma_{k|t}(L_i^p)'\big]^{-1/2}\big(b_{i,k}^p-L_i^pJ_{k|t}^x\Sigma_{22.1}(e_j'J_{t+1|t}^x+e_i'J_{k|t}^x)'-L_i^p\hat{\mu}_{k|t,k}\big)\Big).\nonumber
\end{eqnarray}
Similarly, for $j=1,\dots,n_x$, $j$--th component of the first expectation of equation \eqref{07075}, is obtained by the following equation
\begin{eqnarray}\label{07078}
&&\mathbb{\hat{E}}_{t,k}\Big[\exp{\{\tilde{x}_{j,t+1}\}}1_{E_{i,k}^p}\Big|\mathcal{H}_t\Big]=\exp\bigg\{e_j'\hat{\mu}_{t+1|t,k}+\frac{1}{2}e_j'\Sigma_{t+1|t}e_j\bigg\}\\
&&\times\bar{\Phi}_{n_x}\Big(\big[L_i^p\Sigma_{k|t}(L_i^p)'\big]^{-1/2}\big(b_{i,k}^p-L_i^pJ_{k|t}^x\Sigma_{22.1}(J_{t+1|t}^x)'e_j-L_i^p\hat{\mu}_{k|t,k}\big)\Big).\nonumber
\end{eqnarray}

For the unit--liked life insurance products, according to $(t,k)$--forward probability measure, for $k=t+1,\dots,T$, we have that
\begin{eqnarray}\label{07079}
&&U_{k|t}(\mathcal{H}_t):=\mathbb{\tilde{E}}\big[D_k\big((M_k-G_k)^++G_k\big)\overline{X}_{t+1}\big|\mathcal{H}_t\big]=D_tD_{t+1}B_{t,k}(\mathcal{H}_t)\\
&&\times\sum_{i=1}^{n_x}\bigg(w_{i,k}\mathbb{\hat{E}}_{t,k}\Big[\exp{\{\tilde{x}_{t+1}+\tilde{x}_{i,k}i_{n_x}\}}1_{E_{i,k}^c}\Big|\mathcal{H}_t\Big]-G_k\mathbb{\hat{E}}_{t,k}\Big[\exp{\{\tilde{x}_{t+1}\}}\Big(1_{E_{i,k}^c}-1\Big)\Big|\mathcal{H}_t\Big]\bigg)\nonumber
\end{eqnarray}
Similarly to the segregated funds, for $j=1,\dots,n_x$, the $j$--th components of the first and last expectations of the above equation are found by the following equations
\begin{eqnarray}\label{07080}
&&\mathbb{\hat{E}}_{t,k}\Big[\exp{\{\tilde{x}_{j,t+1}+\tilde{x}_{i,k}\}}1_{E_{i,k}^c}\Big|\mathcal{H}_t\Big]\nonumber\\
&&=\exp\bigg\{e_j'\hat{\mu}_{t+1|t,k}+e_i'\hat{\mu}_{k|t,k}+\frac{1}{2}e_j'\Sigma_{t+1|t}e_j+\frac{1}{2}e_i'\Sigma_{k|t}e_i+e_j'J_{t+1|t}^x\Sigma_{22.1}(J_{k|t}^x)'e_i\bigg\}\\
&&\times\bar{\Phi}_{n_x}\Big(\big[L_i^c\Sigma_{k|t}(L_i^c)'\big]^{-1/2}\big(b_{i,k}^c-L_i^cJ_{k|t}^x\Sigma_{22.1}(e_j'J_{t+1|t}^x+e_i'J_{k|t}^x)'-L_i^c\hat{\mu}_{k|t,k}\big)\Big)\nonumber
\end{eqnarray}
and
\begin{eqnarray}\label{07081}
&&\mathbb{\hat{E}}_{t,k}\Big[\exp{\{\tilde{x}_{j,t+1}\}}\Big(1_{E_{i,k}^c}-1\Big)\Big|\mathcal{H}_t\Big]=\exp\bigg\{e_j'\hat{\mu}_{t+1|t,k}+\frac{1}{2}e_j'\Sigma_{t+1|t}e_j\bigg\}\\
&&\times\bigg(\bar{\Phi}_{n_x}\Big(\big[L_i^c\Sigma_{k|t}(L_i^c)'\big]^{-1/2}\big(b_{i,k}^c-L_i^cJ_{k|t}^x\Sigma_{22.1}(J_{t+1|t}^x)'e_j-L_i^c\hat{\mu}_{k|t,k}\big)\Big)-1\bigg).\nonumber
\end{eqnarray}
By substituting equations \eqref{07077}, \eqref{07078}, \eqref{07080}, and \eqref{07081} into equations \eqref{07075} and \eqref{07079}, one obtains conditional expectations, which are given in equations \eqref{07073} and \eqref{07074}. Therefore, due to equations \eqref{07060}--\eqref{07063}, \eqref{07073}, and \eqref{07074}, and the tower property of conditional expectation, we get $\Lambda_{t+1}$s of the equity--linked life insurances on maximum of asset prices:

\begin{itemize}
\item[1.] for the $T$--year term life insurance that corresponds to segregated fund on maximum of weighted prices of the domestic assets, we have
\begin{equation}\label{07082}
\Lambda_{t+1}=1_{\{T_x>t\}}\sum_{k=t}^{T-1}\bigg\{\sum_{s}\int_{\Pi_{\hat{s}},\Gamma_{\hat{s}}}S_{k+1|t}(\mathcal{H}_t) \tilde{f}(\Pi_{\hat{s}},\Gamma_{\hat{s}},s|\mathcal{F}_t)d\Pi_{\hat{s}} d\Gamma_{\hat{s}} \bigg\}{}_{k-t}p_{x+t}q_{x+k}-\lcterm{\overline{S}}{x+t}{T-t}\overline{X}_{t}
\end{equation}
where $\lcterm{\overline{S}}{x+t}{T-t}:=D_t\lcterm{S}{x+t}{T-t}$ is a discounted net single premium of $T$--year term life insurance, corresponding to segregated fund, see equation \eqref{07060};

\item[2.] for the $T$--year pure endowment insurance that corresponds to segregated fund on maximum of weighted prices of the domestic assets, we have
\begin{eqnarray}\label{07083}
\Lambda_{t+1}=1_{\{T_x>t\}}\bigg\{\sum_{s}\int_{\Pi_{\hat{s}},\Gamma_{\hat{s}}}S_{T|t}(\mathcal{H}_t)\tilde{f}(\Pi_{\hat{s}},\Gamma_{\hat{s}},s|\mathcal{F}_t)d\Pi_{\hat{s}} d\Gamma_{\hat{s}}\bigg\}{}_{T-t}p_{x+t}-\lcend{\overline{S}}{x+t}{T-t}\overline{X}_{t}
\end{eqnarray}
where $\lcend{\overline{S}}{x+t}{T-t}:=D_t\lcend{S}{x+t}{T-t}$ is a discounted net single premium of $T$--year pure endowment insurance, corresponding to segregated fund, see equation \eqref{07061};

\item[3.] for the $T$--year term life insurance that corresponds to unit--linked life insurance on a maximum of weighted prices of the domestic assets, we have
\begin{equation}\label{07084}
\Lambda_{t+1}=1_{\{T_x>t\}}\sum_{k=t}^{T-1}\bigg\{\sum_{s}\int_{\Pi_{\hat{s}},\Gamma_{\hat{s}}}U_{k+1|t}(\mathcal{H}_t)\tilde{f}(\Pi_{\hat{s}},\Gamma_{\hat{s}},s|\mathcal{F}_t)d\Pi_{\hat{s}} d\Gamma_{\hat{s}}\bigg\}{}_{k-t}p_{x+t}q_{x+k}-\lcterm{\overline{U}}{x+t}{T-t}\overline{X}_{t}
\end{equation}
where $\lcterm{\overline{U}}{x+t}{T-t}:=D_t\lcterm{U}{x+t}{T-t}$ is a discounted net single premium of $T$--year term life insurance, corresponding to unit--linked life insurance, see equation \eqref{07062};

\item[4.] for the $T$--year term life insurance that corresponds to unit--linked life insurance on a maximum of weighted prices of the domestic assets, we have
\begin{equation}\label{07085}
\Lambda_{t+1}=1_{\{T_x>t\}}\bigg\{\sum_s\int_{\Pi_{\hat{s}},\Gamma_{\hat{s}}}U_{T|t}(\mathcal{H}_t)\tilde{f}(\Pi_{\hat{s}},\Gamma_{\hat{s}},s|\mathcal{F}_t)d\Pi_{\hat{s}} d\Gamma_{\hat{s}}\bigg\}{}_{T-t}p_{x+t}-\lcend{\overline{U}}{x+t}{T-t}\overline{X}_{t}
\end{equation}
where $\lcend{\overline{U}}{x+t}{T-t}:=D_t\lcend{U}{x+t}{T-t}$ is a discounted net single premium of $T$--year pure endowment insurance, corresponding to unit--linked life insurance, see equation \eqref{07063}.
\end{itemize}
As a result, if we substitute equations \eqref{07072} and \eqref{07082}--\eqref{07085} into equation \eqref{07065}, we can obtain locally risk--minimizing strategies for the equity--linked life insurance products on a maximum of asset prices.

\section{Conclusion}

Economic variables play important roles in any economic model, and sudden and dramatic changes exist in the financial market and economy. The VAR process is workhorse for economic analysis and forecast, but it entails a risk of over--parametrization. Therefore, in the paper, we introduce the Bayesian MS--VAR process and obtain pricing and hedging formulas for the segregated funds and unit--linked life insurance products on a maximum of several asset prices using the risk--neutral valuation method and locally risk--minimizing strategy. We assume that the regime--switching process is generated by a homogeneous Markov process and the residual process follows a heteroscedastic model, where we suppose that conditional covarince matrix of the residual process does not depend on the residual process. 

It should be noted that Bayesian MS--VAR process contains a simple VAR process, vector error correction model (VECM), BVAR, and MS--VAR process. To use our proposed model, which is based on Bayesian MS--VAR process, one may apply Monte--Carlo methods. The early Monte--Carlo methods can be found in \citeA{Krolzig97}, while recent new method, which removes duplication in the regime--switching vector can be found in \citeA{Battulga24g}. A Monte--Carlo method for large BVAR process, we refer to \citeA{Banbura10}. For simple MS--VAR process, maximum likelihood methods are provided by \citeA{Hamilton89,Hamilton90,Hamilton94} and \citeA{Krolzig97}. To summarize, the main advantages of the paper are

\begin{itemize}
\item because we consider VAR process, the spot rate is not constant and is explained by its own and other variables' lagged values,
\item it introduced economic variables, regime--switching, and heteroscedasticity to the equity--linked life insurance products on a maximum of asset prices,
\item it introduced the Bayesian method for valuation of the equity--linked life insurance products on a maximum of asset prices, so the model will overcome over--parametrization,
\item valuation and hedging of the equity--linked life insurance products on a maximum of asset prices is not complicated,
\item and the model contains simple VAR, VECM, BVAR, and MS--VAR processes.
\end{itemize}

\bibliographystyle{apacite}
\bibliography{References}

\end{document}